\documentclass[%
 reprint,
 amsmath,amssymb,
 aip,
]{revtex4-1}

\usepackage{graphicx}
\usepackage{subcaption}
\usepackage{dcolumn}
\usepackage{bm}

\usepackage{xcolor}

\newcommand{\secref}{Section~\ref}
\newcommand{\appref}{Appendix~\ref}
\newcommand{\figref}[2][{}]{\figurename~\ref{#2}#1}

\newcommand*\dif{\mathop{}\!\mathrm{d}}

\begin{document}


\title{Stochastic effects on phase-space holes and clumps in kinetic systems near marginal stability}

\author{B. J. Q. Woods}
\email{benjamin.woods@york.ac.uk}
\affiliation{%
   York Plasma Institute, Department of Physics, University of York, Heslington, York, YO10 5DD, UK
}%
\author{V. N. Duarte}
\email{vduarte@pppl.gov}
\affiliation{%
   Princeton Plasma Physics Laboratory, Princeton University, Princeton, NJ 08543, USA
}%
\affiliation{%
   Institute of Physics, University of S\~{a}o Paulo, S\~{a}o Paulo, SP 05508-090, Brazil
}%
\author{A. P. De-Gol}
\affiliation{%
   York Plasma Institute, Department of Physics, University of York, Heslington, York, YO10 5DD, UK
}%
\author{N. N. Gorelenkov}
\affiliation{%
   Princeton Plasma Physics Laboratory, Princeton University, Princeton, NJ 08543, USA
}%
\author{R. G. L. Vann}
\affiliation{%
   York Plasma Institute, Department of Physics, University of York, Heslington, York, YO10 5DD, UK
}%

\date{\today}

\begin{abstract}
The creation and subsequent evolution of marginally-unstable modes have been observed in a wide range of fusion devices. This behaviour has been successfully explained, for a single frequency shifting mode, in terms of phase-space structures known as a `hole' and `clump'.

Here, we introduce stochasticity into a 1D kinetic model, affecting the formation and evolution of resonant modes in the system. We find that noise in the fast particle distribution or electric field leads to a shift in the asymptotic behaviour of a chirping resonant mode; this noise heuristically maps onto microturbulence via canonical toroidal momentum scattering, affecting hole and clump formation. The profile of a single bursting event in mode amplitude is shown to be stochastic, with small changes in initial conditions affecting the lifetime of a hole and clump. As an extension to the work of Lang and Fu[\onlinecite{lang2011nonlinear}], we find that an intermediate regime exists where noise serves to decrease the effective collisionality, where microturbulence works against pitch-angle scattering.
\end{abstract}

\maketitle



\section{\label{sec:int}Introduction}
Toroidal Alfv\'{e}n eigenmodes, (TAEs) are of particular interest to the fusion community. With frequencies on the order of 100 kHz [\onlinecite{heidbrink2008basic}], they have the ability to become amplified via RF heating and energetic particles. Due to defects in the magnetic field periodicity in real tokamaks, spatially localised modes (gap TAEs) can exist in the frequency gap between TAEs and BAEs (beta-induced Alfv\'{e}n eigenmodes; frequency $<$ 50 kHz, close to GAM frequency). Unlike continuum Alfv\'{e}n modes, gap TAEs exist as coherent waveforms which are resilient to shear damping. Energetic particles undergoing Landau resonance with plasma waves under go radial diffusion, which can lead to large fast particle losses in tokamaks [\onlinecite{pinches2004role,fredrickson2006fast,heidbrink2008basic,gorelenkov2014energetic}]; gap TAEs exist for longer timescales than continuum modes, allowing for greater particle loss.

It is well known that hole and clump structures [\onlinecite{berk1997spontaneous,breizman1997critical,wang2013hole}] can form in the non-linear phase of the evolution of an energetic particle driven mode, such as a TAE. It is understood that mode chirping is directly correlated with hole and clump structures; consequently, chirping modes can allow for greater radial diffusion. As a result, even in the case of continuum Alfv\'{e}n eigenmodes, rapid mode chirping can also lead to a significant channel for fast ion loss; in such a case, the rate of energy loss via chirping is comparable to the sum of damping rates (e.g. collisional, radiative, continuum damping).

\begin{figure}[h!]
\includegraphics[width=0.35\textwidth]{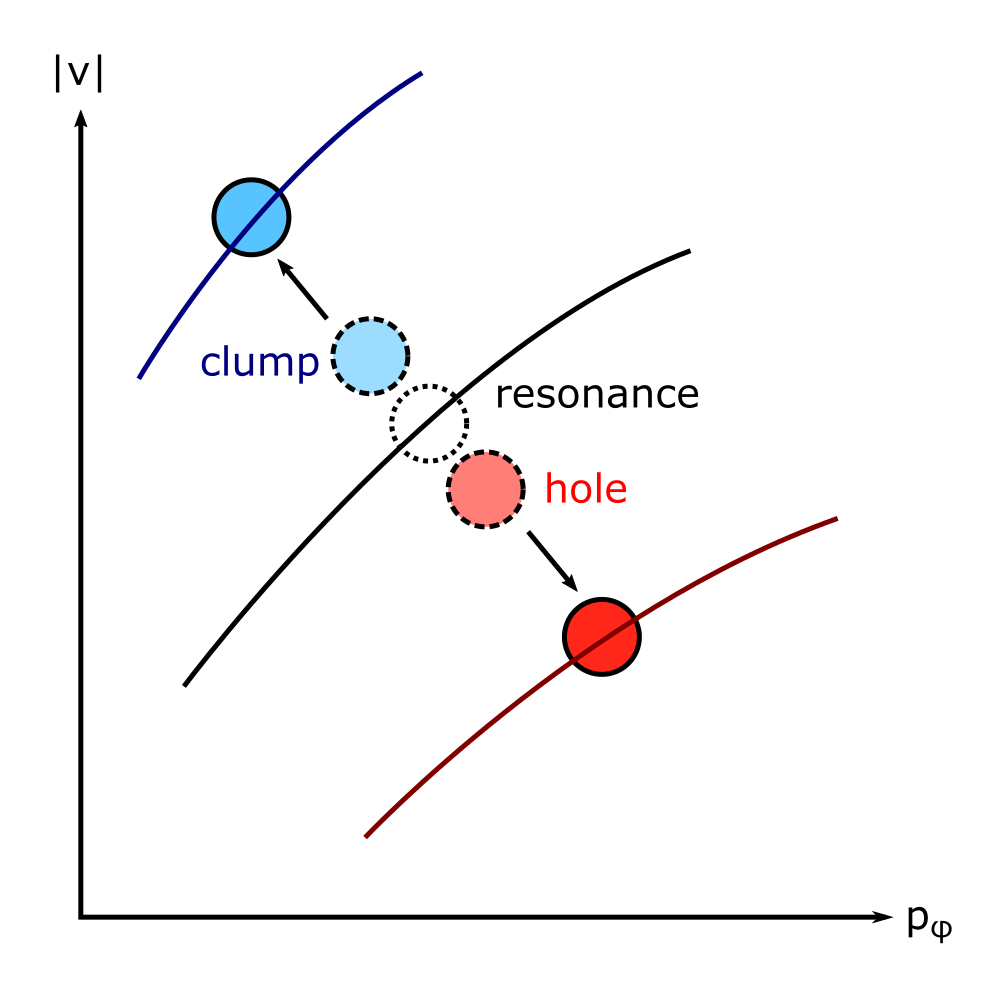}
\caption{A sketch of a hole and clump on a distribution function peaked near the origin for a chirping, resonant mode interacting with a tokamak plasma. The resonance $\omega + (m + l) \omega_{\theta}- n \omega_{\varphi} = 0$ undergoes pitchfork bifurcation as the hole and clump move. Stochastic fluctuations modelled in this paper heuristically map onto fluctuations a function of $p_{\varphi}$; these effect the formation and evolution of the hole and clump.}
\label{fig:res}
\end{figure}

The effects of random, small-scale phenomena have been previously examined in the literature[\onlinecite{berk1997spontaneous,breizman1997critical,wang2013hole,gorelenkov2014energetic}]: mechanisms such as pitch-angle scattering can destroy holes and clumps. However, the effect of microturbulence on the evolution of a bursting mode in its non-linear phase is relatively unexplored. Recent work by Duarte \emph{et al.} [\onlinecite{duarte2017prediction, duarte17-2}] proposes that enhanced stochasticity in resonant particle dynamics, in the form of fast ion microturbulence, can be a key mechanism for chirping suppression in several tokamak scenarios. The prediction stimulated dedicated experiments on DIII-D by Van Zeeland \emph{et al.} [\onlinecite{vanzeeland17}] with negative triangularity, known for suppressing drift-like instabilities. The experiments have shown a clear correlation between chirping emergence and scenarios with very low turbulent activity.

This motivates the work detailed in this paper: here, we carefully explore the inclusion of stochasticity into 1D kinetic models, allowing us to examine in closer detail the resultant effects on the evolution of resonant modes.

\section{\label{sec:comp}Stochastic model}

\subsection{Resonant damping}f
In a tokamak, TAEs resonate with a quasi-2D fast-ion distribution function where the linear stability is determined by competing $\text{d}f/\text{d}|v|$ and $\text{d}f/\text{d}p_{\varphi}$ [\onlinecite{heidbrink2008basic}]. These correspond to resonance with the poloidal and toroidal transit frequencies $\omega_{\theta}$ and $\omega_{\varphi}$ respectively, [\onlinecite{heidbrink2008basic,breizman2010nonlinear}] given by $\omega + (m + l) \omega_{\theta} - n \omega_{\varphi} = 0$. Here, $m$ and $n$ are the poloidal and toroidal modenumbers, and $l \in \mathbb{Z}^*$ correspond to poloidal harmonics of the drift velocity (see \figref{fig:res}).

We model the same key instability physics by examining hole and clump formation on a 1D bump-on-tail distribution function. This allows us to model energetic particle drive via the positive slope of the distribution function between the bulk and the beam, while modelling energetic losses as a damping term. [\onlinecite{berk1997spontaneous,breizman1997critical,berk1995numerical,vann2003fully,degol10}]

The evolution of the system is determined by coupling the Boltzmann equation to Maxwell's equations; our model is given here by a multiple species generalisation of kinetic models used by Vann et al.[\onlinecite{vann2003fully}] and De-Gol [\onlinecite{degol10}]:

\begin{subequations}
\begin{align}
\label{eq:xv1}
\partial_t f_l &= C_l[f_l] - v \partial_x f_l - \dfrac{q_l}{m_l}E \partial_v f_l & \forall l\\
\label{eq:xv2}
\partial_t E &= - \dfrac{1}{\epsilon_0} \displaystyle\int\limits_{-\infty}^{\infty} v \sum\limits_l (q_l f_l) \dif  v - \alpha E
\end{align}
\end{subequations}

where $\{C_l\}$ are collision operators, $f_l(x,v,t)$ is the fast particle distribution function for the $l^{\text{th}}$ species, $E(x,t)$ is the electric field, and $\epsilon_0$ is the permittivity of free space. Damping is effected in the system via $\alpha(x,t)$, and acts as a  sink of electromagnetic field energy. Formally, one can show that this augmentation still preserves momentum and energy globally (see derivation in \appref{app:bbm} from classical field theory).

\subsection{\label{sec:twospecies} Two species model for turbulence}
Here, we model the plasma by using two separate distributions of identical particles: a fast ion distribution $f_{\text{ion}}$ with a deterministic trajectory, and a turbulent distribution $f_{\text{tur}}$ with a stochastic trajectory. By modelling the plasma using two separate distributions, one is able to vary with ease the fraction of particles that are turbulent. Fluctuations in $f_{\text{tur}}$ lead to fluctuations in the electric field via \eqref{eq:xv2}. Fluctuations in the electric field interact with $f_{\text{ion}}$ via \eqref{eq:xv1}, heuristically mapping via $\alpha E$ onto the energy exchange associated with particle resonance along $\dif f/\dif |v|$. Accordingly, the model captures stochasticity in a comparable way to momentum scattering in $p_{\varphi}$ as previously explored by Lang and Fu.[\onlinecite{lang2011nonlinear}]

We treat the electrons as existing within a neutralising background that do not move with the ions, or resonate with the electric field. This allows the model to become a two species plasma:

\begin{subequations}
\begin{align}
\partial_t f_{\text{ion}} &= C_{\text{ion}}[f_{\text{ion}}] - v \partial_x f_{\text{ion}} - E \partial_v f_{\text{ion}} \\
\partial_t f_{\text{tur}} &= C_{\text{tur}}[f_{\text{tur}}] - v \partial_x f_{\text{tur}} - E \partial_v f_{\text{tur}} \\
\partial_t E &= - \displaystyle\int\limits_{-\infty}^{\infty}v(f_{\text{ion}} + f_{\text{tur}}) \dif  v - \alpha E
\end{align}
\end{subequations}

The normalisation is as follows:

\begin{equation}
\begin{array}{r c c c c c l}
N_{\text{ion}} = 1 &;& m_{\text{ion}} = 1 &;& \omega_p = 1 &;& v_{\text{th}} = 1
\end{array}
\end{equation}

where for the non-turbulent fast ions, $N_{\text{ion}}$ is the number of particles, $\omega_p$ is the plasma frequency. The quasi-thermal quantity $v_{\text{th}}$ normalises the energy of the system, but the equilibrium is not a thermal equilibrium. We utilise a Fokker-Planck diffusive collision operator for $C_{\text{tur}}$, in alignment with kinetic descriptions of mode chirping based on pitch angle scattering $\nu_{\text{eff}}$ in the literature [\onlinecite{breizman1997critical, lang2010nonlinear, duarte2017prediction, duarte17-2}]; this also allows for comparison with the LBQ[\onlinecite{ghantous2014comparing}] and BOT[\onlinecite{lilley2009destabilizing}] codes. However, it is important to note that the operator relaxes to the initial distribution function; accordingly, energy transfer may occur during relaxation.

For $C = \mathcal{O}(\partial_v^3)$, one can show that the resultant hyperjerk equation $\partial_t f = C[f]$ at fixed $t$ can be represented in the form $\partial_v \mathbf{g} = \mathbf{F}(\mathbf{g})$ where $\mathbf{g} \in \mathbb{R}^n$ where $n \geq 3$, and $\mathbf{F}(\mathbf{g})$ is a smooth function [\onlinecite{linz1997nonlinear}]. Via the Poincar\'{e}-Bendixson theorem [\onlinecite{poincare1885courbes, bendixson1901courbes, chlouverakis2006chaotic}], this meets the minimum requirements for chaotic behaviour; a simple \emph{reductio ad absurdum} shows that replacing $n$ with a value less than 3 prevents the formation of chaotic solutions.

Consequently, we find that the collision operator $C_{\text{tur}}$ must include terms $\mathcal{O}(\partial_v^3)$ to allow for stochastic behaviour. In lieu of an implemented collisional model beyond Fokker-Planck theory, we instead define the trajectory of $f_{\text{tur}}$ \emph{ad hoc} to investigate the key resultant physics.

As a result:

\begin{subequations}
\begin{align}
\partial_t f_{\text{ion}} &= \nu \partial_v^2 (f_{\text{ion}} - F_0) - v \partial_x f_{\text{ion}} - E \partial_v f_{\text{ion}} \\
f_{\text{tur}} &= f_{\text{tur}}(x,v,t) \\
\partial_t E &= - \displaystyle\int\limits_{-\infty}^{\infty} v(f_{\text{ion}} + f_{\text{tur}}) \dif  v - \alpha E
\end{align}
\end{subequations}

where $F_0(v) \equiv f_{\text{ion}}(x,v,t=0)$.

\onecolumngrid

\begin{figure}
\centering
	\begin{subfigure}[t]{0.4\textwidth}
		\includegraphics[width=0.8\textwidth]{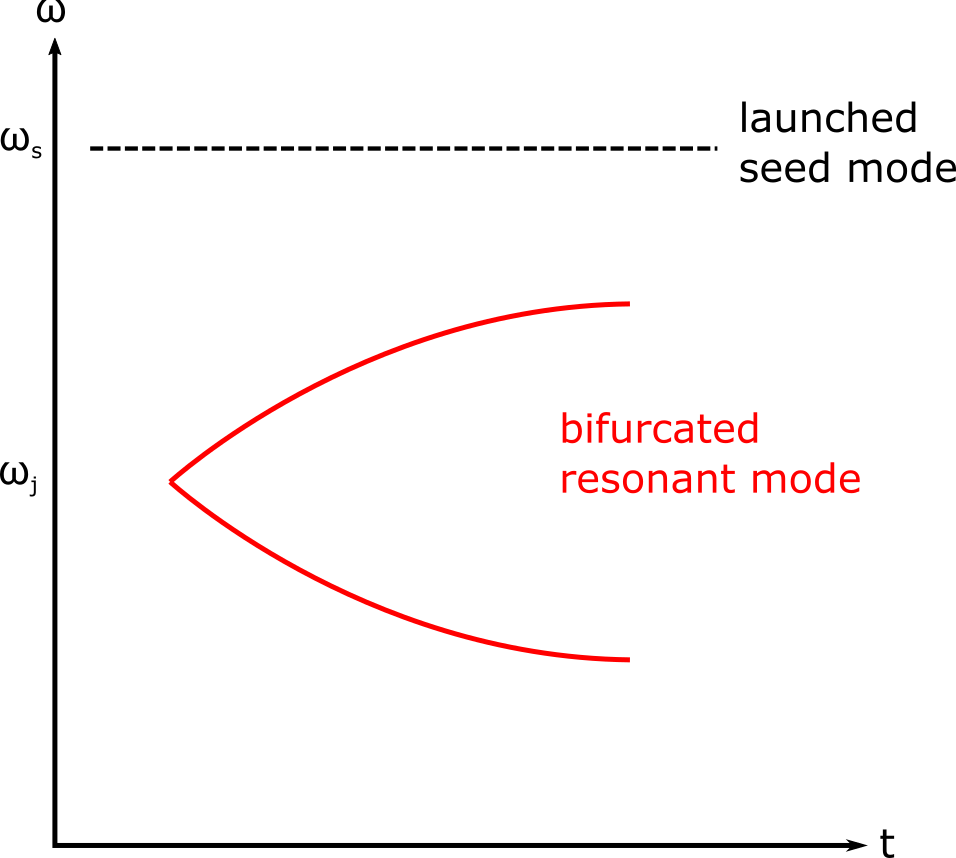}
		\subcaption{Sketch of a frequency spectrogram for a seed mode ($\omega_s$) launched at a frequency much greater than that of a resonant mode ($\omega_j(t=0) \sim \omega_p$).}
		\label{fig:seed}
	\end{subfigure}
	\hspace{40px}
	\begin{subfigure}[t]{0.4\textwidth}
		\includegraphics[width=0.8\textwidth]{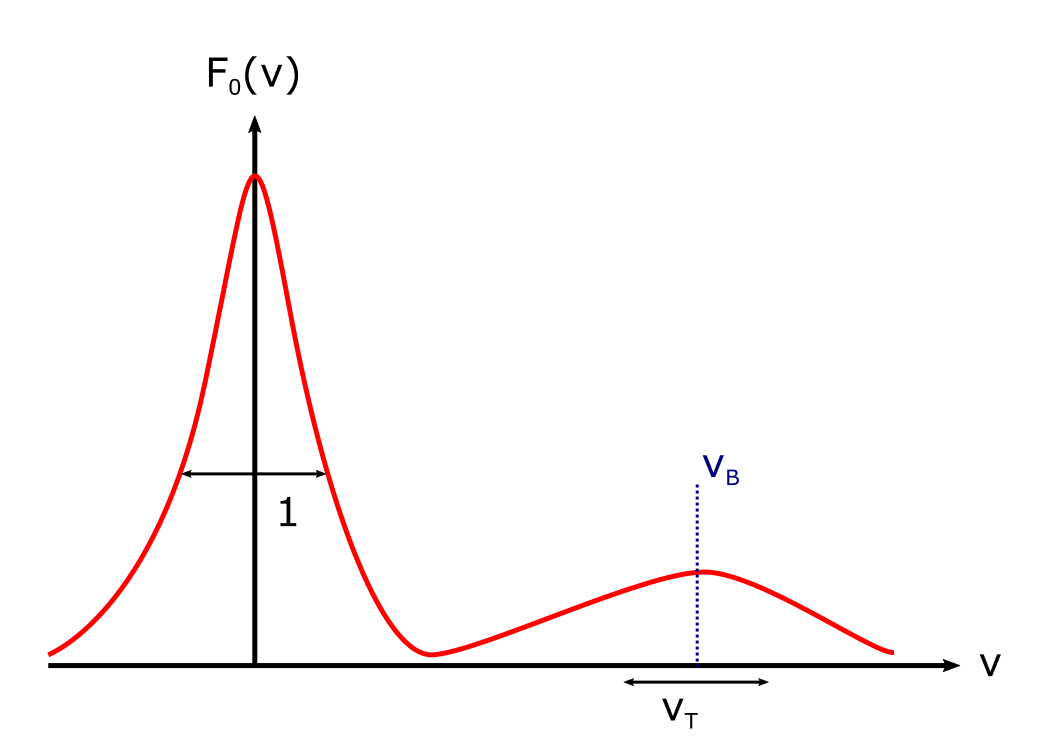}
		\subcaption{Sketch of the bump-on-tail distribution, $F_0(v)$.}
		\label{fig:bot}
	\end{subfigure}
\caption{Diagrams detailing non-stochastic drive for the system. A resonant mode undergoing chirping does so in a frequency bandwidth much lower than the frequency of the launched seed mode (\figref{fig:seed}), with mode drive generated via inverse Landau damping on the positive slope of $F_0(v)$ (\figref{fig:bot}) between 0 and $v_B$.}
\label{fig:drive}
\end{figure}

\twocolumngrid

\subsection{\label{sec:approx} Approximations}
We make a few assumptions to simplify the model computationally. These rely on a Fourier series representation of the distribution function and other quantities:

\begin{equation}
f(x,v,t) = \dfrac{1}{2} \sum\limits_j \left[ f_j(v,t) e^{ik_j x} + f_j^*(v,t) e^{-ik_j x} \right]
\end{equation}

where we sum over a set of initial modes, $\{j\}$. We define the velocity Fourier transform by symmetric definitions:

\begin{equation}
\begin{array}{r l}
\tilde{f}(x,s,t) &= \dfrac{1}{\sqrt{2 \pi}} \displaystyle\int\limits_{-\infty}^{\infty} f(x,v,t) e^{-isv} \dif v \\
f(x,v,t) &= \dfrac{1}{\sqrt{2 \pi}} \displaystyle\int\limits_{-\infty}^{\infty} \tilde{f}(x,s,t) e^{isv} \dif v 
\end{array}
\end{equation}

First, we replace $\alpha(x) E(x)$ with $\alpha(x) \ast E(x)$; via the convolution theorem, this allows for a piecewise product of $\alpha$ and $E$ in k-space. However, as it does not preserve the canonical form of the Hamiltonian (see \appref{app:bbm}), conservation of energy is violated except for the trivial case where $\alpha(x)$ is a constant. Here, we examine this trivial case, and therefore $\forall j: \alpha_j = 2\alpha $.

Secondly, we ignore three-wave coupling; this means that no modes exist except for harmonics of the initial set of modes. We allow this as three-wave coupling of gap TAEs will generate modes which exist in the Alfven continuum, which are quickly dissipated. [\onlinecite{heidbrink2008basic}]

Thirdly, we also ignore all harmonics of the initial set of modes except for the fundamental; this is justified by requiring physically that these harmonics are rapidly damped.

Finally, we force $E_0$ to be evanescent, and set it to 0 via boundary conditions (see \appref{app:canon}), removing mean current from the Maxwell-Ampere law. This leads to the caveat:

\begin{equation}
\label{eq:caveat}
\bigg|\alpha_0 E_0 \bigg|  >> \bigg| \displaystyle\int\limits_{-\infty}^{\infty} v(f_{\text{ion},0} + f_{\text{tur},0}) \dif v \bigg|
\end{equation}

In principle, \eqref{eq:caveat} is violated, but we find that the induced mean current is very small. Overall, for a single mode simulation after spectral decomposition and velocity Fourier transforms:

\begin{subequations}
\begin{align}
\label{eq:ks1}
&\partial_t \tilde{f}_{\text{ion},j} = -\nu s^2 \tilde{f}_{\text{ion},j} + k_j \partial_s \tilde{f}_{\text{ion},j} - E_j \tilde{f}_{\text{ion},0} \\
\label{eq:ks2}
&\begin{array}{r l}
\partial_t \tilde{f}_{\text{ion},0} &= -\nu s^2 (\tilde{f}_{\text{ion},0} - F_0) \\
&\hspace{20pt} - \dfrac{is}{4} \left[ E_j^* \tilde{f}_{\text{ion},j} + \text{c.c.} \right] 
\end{array} \\
\label{eq:ks3}
&\tilde{f}_{\text{tur}} = \tilde{f}_{\text{tur}}(x,s.t) \\
\label{eq:ks4}
&\partial_t E_j = - \displaystyle\int\limits_{-\infty}^{\infty} \left[W (\tilde{f}_{\text{ion},j} + \tilde{f}_{\text{tur},j}) \right]\dif  s - \dfrac{1}{2} \alpha_j E_j \\
\label{eq:ks5}
&E_0 = 0
\end{align}
\end{subequations}

where $W(s) \in \mathbb{I}$ is the formally divergent integral (which acts a removable singularity in the current):

\[
W(s) \equiv \dfrac{1}{\sqrt{2 \pi}}\int\limits_{-\infty}^{\infty} v e^{isv} \dif v
\]

\clearpage

\onecolumngrid

\begin{figure}
\centering
	\begin{subfigure}[t]{0.4\textwidth}
		\includegraphics[width=0.7\textwidth]{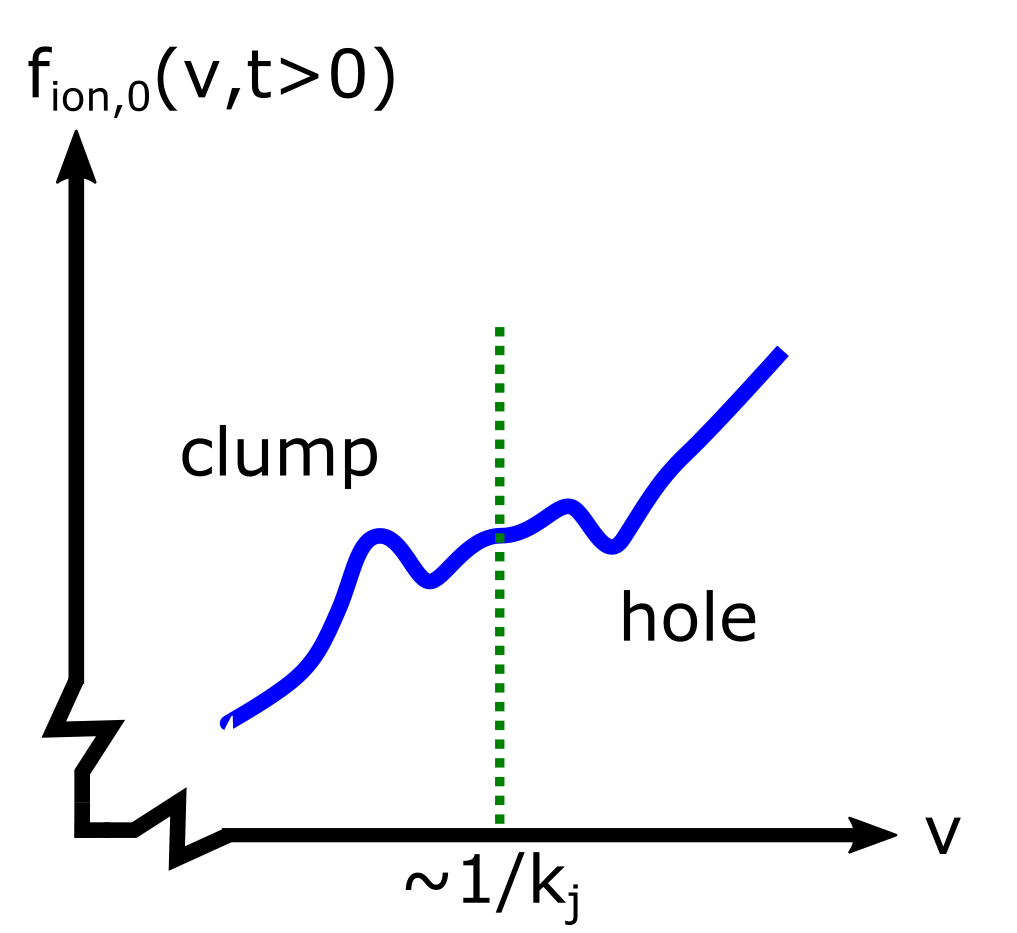}
		\subcaption{A sketch of a typical set of values for the fast ion distribution $f_{\text{ion},0}(v,t>0)$: in the vicinity of $v \sim 1/k_j$, a hole and clump can form during resonance.} 
		\label{fig:hac}
	\end{subfigure}
	\hspace{40px}
	\begin{subfigure}[t]{0.4\textwidth}
		\includegraphics[width=0.7\textwidth]{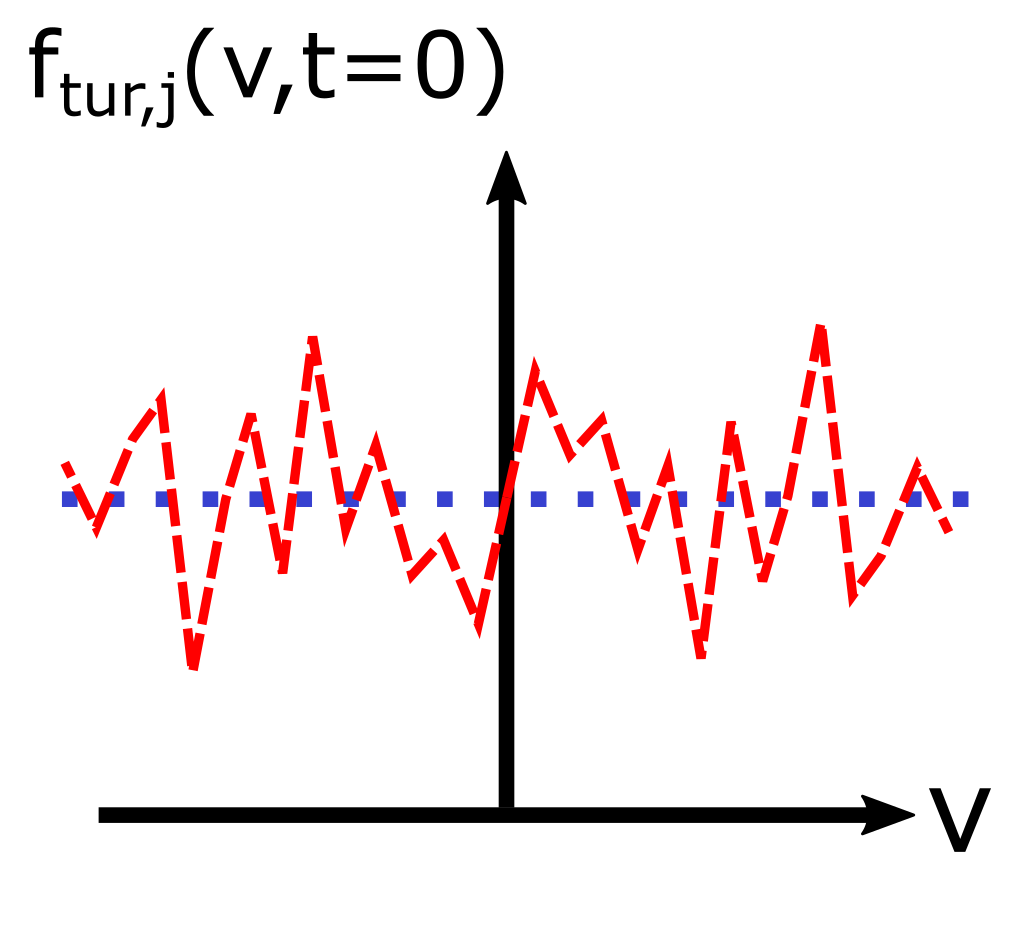}
		\subcaption{A sketch of a typical set of values for the turbulent distribution $f_{\text{tur},j}(v,t=0)$: blue dotted line denotes $\langle f_{\text{tur},j} \rangle_v(v,t=0)$, red dashed line denotes $\mathcal{N}_{f,j}(v,t=0)$.}
		\label{fig:turbf}
	\end{subfigure}
\caption{Sketches highlighting features of $f_{\text{tur}}$ and $f_{\text{ion}}$. Both $f_{\text{tur}}$ and $f_{\text{ion}}$ have a constant number of particles, and $f_{\text{tur}}$ contains no net energy.}
\label{fig:dist}
\end{figure}

\twocolumngrid

\subsection{Seed electric field}
We define $E$ using three parts: a perturbation $\delta E$, a deterministic part $E_{\mathcal{S}}$, and a stochastic part $E_{\mathcal{N}}$:

\begin{equation}
\label{eq:E}
E = \delta E + E_{\mathcal{S}} + E_{\mathcal{N}}
\end{equation}

One can show that \eqref{eq:ks4} now takes the form:

\[
\partial_t \delta E_j = - \int W \cdot (\tilde{f}_{\text{ion},j} + \tilde{f}_{\text{tur},j} ) \dif s - \dfrac{1}{2} \alpha_j \delta E_j + \mathcal{D}_j
\]

where the drive term $\mathcal{D}_j$ is given by:

\[
\mathcal{D}_j \equiv \partial_t (E_{\mathcal{S},j} + E_\mathcal{N}) + \dfrac{1}{2} \alpha_j (E_{\mathcal{S},j} +  E_\mathcal{N})
\]

We desire that the seed terms $E_{\mathcal{S}}$ and $E_{\mathcal{N}}$ exchange no free energy with the system. \emph{Ad hoc} definitions which are independent of $\delta E$ and $\{f_l\}$ achieve this:

\[
\begin{array}{r c l}
E_{\mathcal{S},j} = D(1 - \epsilon_{E,j}) e^{i\omega_s t} &;& E_{\mathcal{N},j} = E_{\mathcal{N},j} (t)
\end{array}
\]

where $\omega_s$ is the seed mode frequency, and $\epsilon_{E,j}$ allows us to define stochastic seeding of the electric field with wavenumber $k_j$; $\epsilon_{E,j} = 1$ is fully stochastic seeding, $\epsilon_{E,j} = 0$ is non-stochastic seeding. $D$ is the typical amplitude of $(E_{\mathcal{S}} + E_{\mathcal{N}})$.

For $E_{\mathcal{N},j} (t)$, we use Gaussian noise, with a mean value of 0, and a domain of $[-\infty, \infty]$. The standard deviation of the values $E_{\mathcal{N},j} (t)$ takes is $\sigma_{E,j}$, which we treat as the typical amplitude. Accordingly, $\sigma_{E,j} = \epsilon_{E,j} D$.

In order for the seed field ($E_{\mathcal{N}} + E_{\mathcal{S}}$) to not directly interact with the plasma (and attenuate), the constituent frequencies must be much greater than the plasma frequency. If we utilise pseudorandom noise, it oscillates on a timescale comparable to the timestep. For $E_{\mathcal{S}}$, we have to enforce $\omega_s >> \omega_p$ (see \figref{fig:seed}).

\subsection{\label{sec:dist} Turbulent distribution}
The fast-ions are modelled using a 1D bump-on-tail distribution (see \figref{fig:bot}):

\begin{equation}
F_0(v) = \dfrac{1}{\sqrt{2 \pi}} \left[ \eta \exp \left( -\dfrac{v^2}{2} \right) + \dfrac{1 - \eta}{v_t} \exp \left( - \dfrac{(v - v_b)^2}{2v_t^2} \right) \right]
\end{equation}

such that $(1 - \eta)$ is the fraction of particles in the beam, $v_t$ is the beam width, and $v_b$ is the beam velocity. We define the turbulent population as the sum of a top-hat function and a noise term; the noise term is defined to yield a particle population of zero (see \figref{fig:turbf}). This allows us to parametrically modify the stochasticity of $f_{\text{tur}}$ via noise without changing the total number of particles. After Fourier transforms:

\begin{subequations}
\begin{align}
\label{eq:f_ion}
\tilde{f}_{\text{ion}}(x,s,t) &= \tilde{F}_0(s) + \tilde{\delta f}_{\text{ion}}(x,s,t) \\
\label{eq:f_turj}
\tilde{f}_{\text{tur,j}}(s,t) &= \dfrac{1}{\sqrt{2 \pi}} \epsilon_{f,j} \text{ sinc } \left(\dfrac{s L_v}{2}\right) + \tilde{\mathcal{N}}_{f,j} \\
\label{eq:f_tur0}
\tilde{f}_{\text{tur,0}}(s,t) &= 0
\end{align}
\end{subequations}

where $\text{sinc}(s) \equiv \sin (s) / s$ is the sinc function, $\mathcal{N}_f (x,s,t)$ is a noise term, and $L_v$ is the length of the codomain of $F_0(v)$. Accordingly, the fraction of non-turbulent particles is $1 / [1 + \epsilon_{f,j}]$. We assume that broadband noise in $v$-space will still be broadband noise in $s$-space, weighted accordingly via Parseval's theorem; accordingly we relate amplitudes as:

\[
\dfrac{|\tilde{\mathcal{N}}_{f,j}|}{|\mathcal{N}_{f,j}|} \approx \dfrac{L_v}{\sqrt{2 \pi}}
\]

For $\mathcal{N}_{f,j}(v,t)$, we use raised cosine noise, with a mean value of 0. The domain is given by:

\[
\mathcal{N}_{f,j}(v,t) \in \left[-\sigma_{f,j} \sqrt{\dfrac{3 \pi^2}{\pi^2 - 6}}, \sigma_{f,j} \sqrt{\dfrac{3 \pi^2}{\pi^2 - 6}} \right]
\]

The typical amplitude is equal to the standard deviation $\sigma_{f,j}$, and accordingly to force positive $f_{\text{tur}}$ everywhere:

\begin{equation}
\begin{array}{r c l}
\sigma_{f,j} \leq \sigma^{(\text{max})}_{f,j} &;& \sigma^{(\text{max})}_{f,j} \equiv \dfrac{\epsilon_{f,j}}{L_v} \sqrt{\dfrac{\pi^2 - 6}{3 \pi^2}}
\end{array}
\end{equation}

We also require for conservation of energy and particle number (at constant $\epsilon_f$) that the $0^{\text{th}}$ and $2^{\text{nd}}$ moment of $f_{\text{tur}}$ vanish. To enable this, we enforce that $\tilde{\mathcal{N}}_f(x,s,t)$ has a real part that is odd, and an imaginary part that is even. One finds that the net energy content of the turbulent distribution is given by:

\begin{equation}
U_{\text{tur}} = \dfrac{1}{24} \epsilon_{f,j} L_v^3
\end{equation}

\section{\label{sec:method}Computational method}
\subsection{DARK}
We utilise a new, modular code based on previous work by Arber, Vann and De-Gol. [\onlinecite{arber2002critical,degol10}]. DARK (\textbf{D}-dimensional \textbf{A}ugmented \textbf{R}esonance \textbf{K}inetic solver) allows for a single framework which can incorporate fundamentally different models and approximations by using a Strang split set of partial flows. Each partial flow is solved separately to yield the full solution across a timestep. \appref{app:strang} details the exact form of the splitting scheme.

Fourier decomposition reduces $N_x$ grid points in $x$ to $(N_k+1)$ equations (here the number of modes $N_k = 1$). By using $s$-space, the code computes collisions and velocity advection at the same time, giving a factor 2 increase in speed, and only requires backwards transforms via FFTW[\onlinecite{fftw}] (\textbf{F}astest \textbf{F}ourier \textbf{T}ransform in the \textbf{W}est) on distribution function output, providing  a potential $\mathcal{O}(N_v \log N_v)$ decrease in computational time on each timestep.

\subsection{Global parameters}
Our global parameters were selected to be:

\begin{equation}
\begin{array}{c}
\begin{array}{r c c c c c l}
v_b = 10 &;& v_t = 4 &;& \eta = 0.95
\end{array} \\
\begin{array}{r c l}
\omega_s = 2.0 &;& k_j = 0.15 
\end{array} \\
\begin{array}{r c c c l}
\Delta t = 0.1 &;& v \in [-28,88] &;& N_v = 8192
\end{array}
\end{array}
\end{equation}

Selecting $k_j = 0.15$ means that if we define $k_j$ as the fundamental eigennumber of the system (the length of the 1D box $L_x = 2 \pi / k_j$), in turn all of the higher harmonics resonate with the bulk particles, where they undergo strong Landau damping. While some studies have shown that strongly Landau damped modes can be non-linearly unstable [\onlinecite{lesur2012nonlinear,lesur2016nonlinear}], here we assume that this is not possible due to small initial mode amplitude; this allows us to justify the lack of three-wave coupling in the model.

The timestep was picked to be small enough to allow for a reasonable frequency analysis without becoming too computationally expensive. At $\Delta t = 0.1$, using a window size of 2000 timesteps, we obtain a frequency resolution $\Delta \omega = \pi / 100$ provided that $\delta \omega < \Delta \omega$ across the timeframe of the bin. Furthermore, the Fourier spectrum of $E_{\mathcal{N}}$ should be dominated by structures in the region $\omega >> \omega_p$. 

It is also a requirement for the code to adhere to the Courant-Friedrichs-Lewy (CFL) [\onlinecite{courant1928partiellen}] limit  for spatial advection in $s$-space via the piecewise parabolic method (PPM)[\onlinecite{colella1984piecewise}] routine; the number of $v$-points $N_v$ and the domain of $v$ adheres to the CFL limit.

Noise in the system is provided by using a pseudonumber random generator (PRNG) with a given seed value.

\subsection{Benchmarking}
\vspace{-20pt}
\begin{figure}[h!]
\centering
\includegraphics[width=0.45\textwidth]{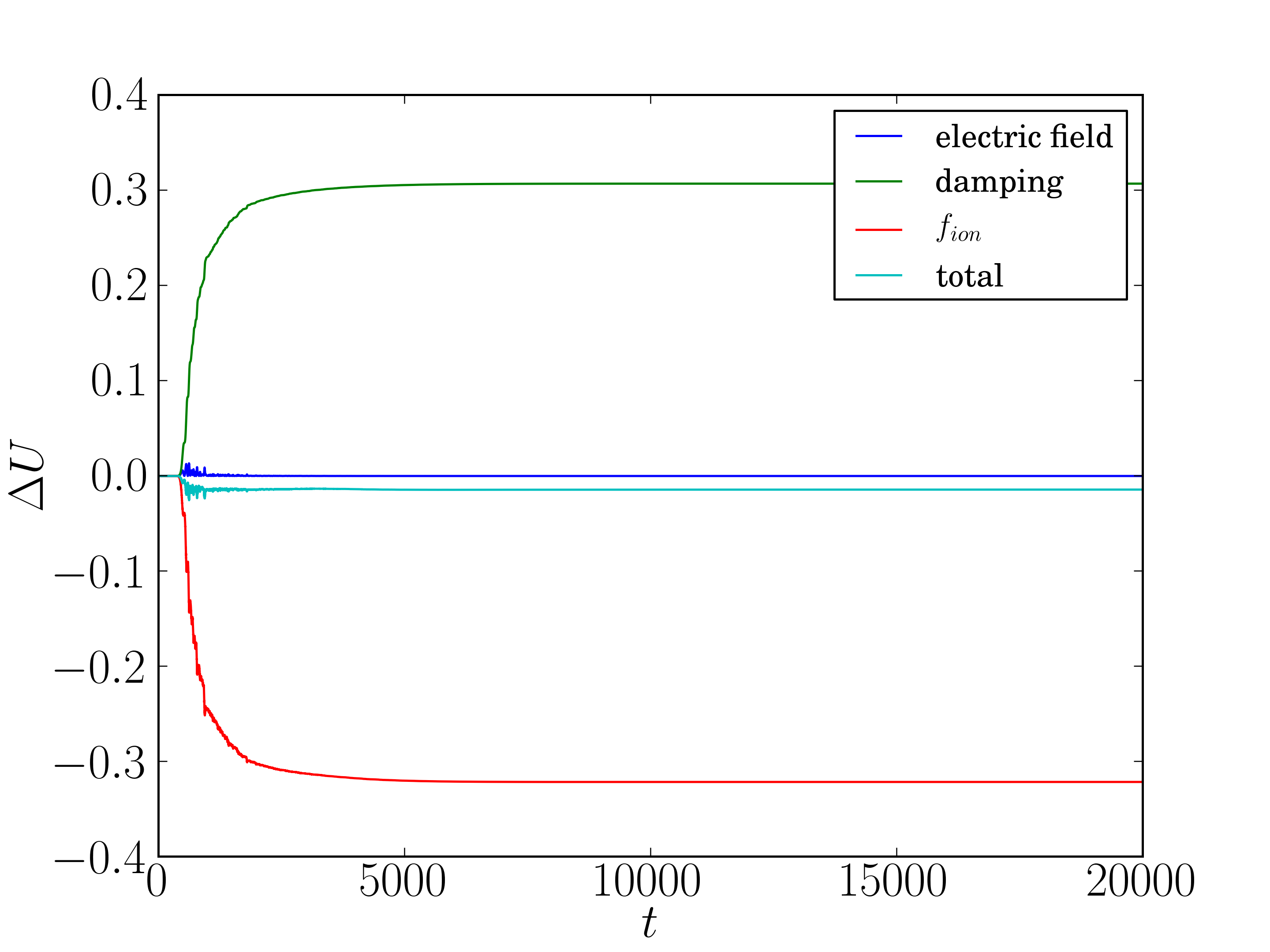}
\caption{Change in energy content ($\Delta U$) for a simulation from \secref{sec:lifetime} with $\alpha_j = 0.6$. The total energy in the system increases linearly in the non-linear phase due to injective collisions.}
\label{fig:conservation}
\end{figure}

Energy is not conserved in these simulations; the Fokker-Planck collisions heat the fast ion distribution, always aiming to restore the energy content in the distribution function to $U_0$:

\begin{equation}
U_0 \equiv \dfrac{1}{2} \int\limits_{-\infty}^{\infty} F_0 v^2 \dif v = \dfrac{1}{2} \left[\eta + (1 - \eta) (v_b^2 + v_T^2)\right]
\end{equation}

For the purpose of benchmarking, we use a very simple model for symmetric mode flattening with a local population transfer $(f_0 - F_0) \sim -(v-v_0) \exp [-(v-v_0)^2]$ near resonance. This yields the corresponding energy flux from collisions in the weakly non-linear regime:

\begin{equation}
\dot{U}_{\text{coll.}} \equiv \dfrac{1}{2} \int \partial_v^2 (f_0 - F_0) v^2 \dif v \sim - v_0 \sqrt{\pi}
\end{equation}

We test the energy conservation by examining a simulation with the same parameters employed in \secref{sec:lifetime}, at $\alpha_j = 0.6$. As is shown in \figref{fig:conservation}, the total energy content in the system is roughly constant initially. As we approach the non-linear phase, the energy lost from $f_{\text{ion}}$ increases sharply. One expects this energy loss to be equal to that lost to the mode $E_j$, and damping $\alpha_j$; the total energy injected into the system via $\nu$ at this time is very small. 

Once the distribution function has suitably relaxed, the deficit in the energy content should asymptote to that lost via damping $\alpha_j$, however we observe a small discrepancy in the energy content (roughly 0.45\% of the total energy content). We believe that this discrepancy is due to approximations made regarding higher harmonics of $f_{\text{ion},j}$ and $E_j$, and deem this discrepancy to be negligible for the single burst simulations examined here.

\section{\label{sec:lifetime} Stochastic lifetime of hole and clump}
\begin{figure}[h!]
\centering
\includegraphics[width=0.35\textwidth]{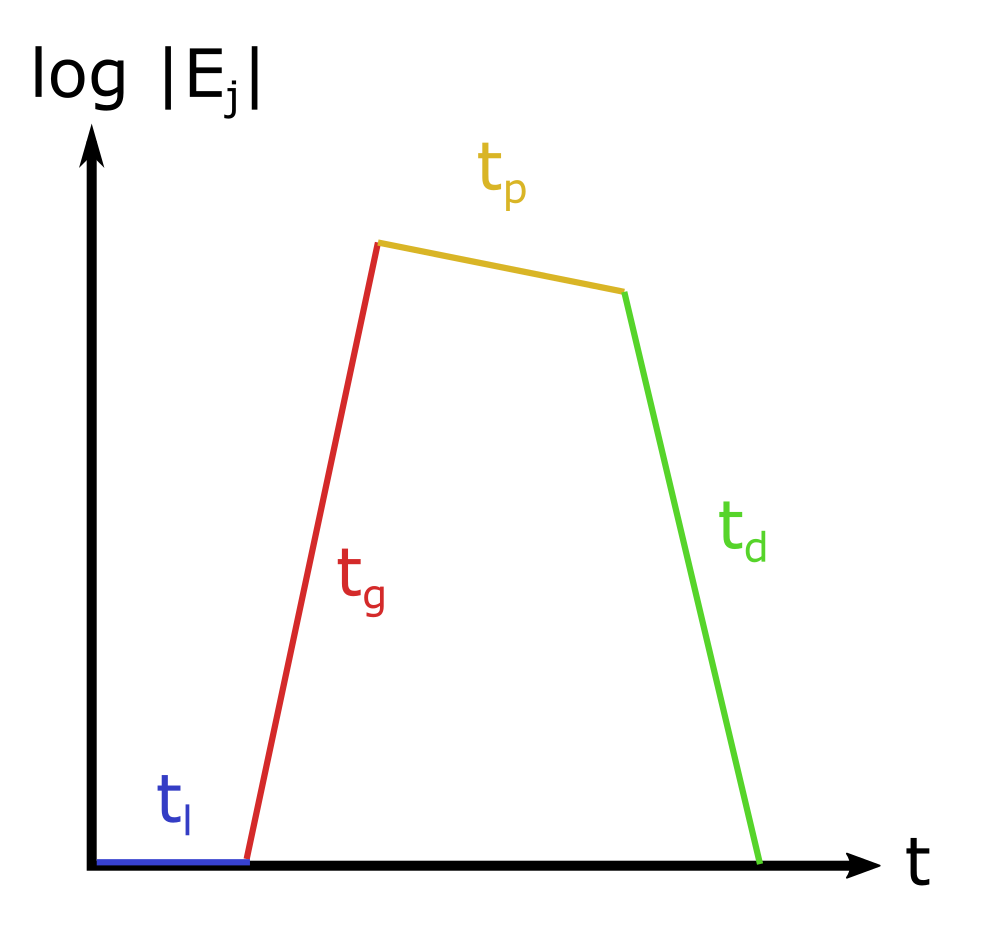}
\caption{A sketch of a bursting mode near marginal stability. Constituent regions $t_l$, $t_g$, $t_p$, and $t_d$ are labelled.}
\label{fig:single}
\end{figure}

Here, we consider a system with noise only in the electric field; that is $\mathcal{N}_f = 0$. We consider no particles in the turbulent population $f_{\text{tur}}$, such that the simulations reduce to single species.

The electric field is only seeded by noise; $D_j = 10^{-7}$ was used, with $\epsilon_E,j = 0$. A set of 2500 simulations were employed, allowing for 50 varying values of $t_{\text{NL}}$, each tested for 50 different initial seeds of the PRNG. $k$ and $\nu$ were fixed to $0.150$ and $10^{-7}$ respectively, with 50 values of $\alpha_j$ on the interval $[0.06,0.158]$. The low value of $\nu$ justifies a relatively large timestep of $\Delta t = 0.1$.

\subsection{Burst characterisation}
To characterise the behaviour of a single bursting event, a set of simulations were used to produce data for the length of four temporal regions: lag, growth, plateau, and decay. Each of these regions are labelled in \figref{fig:single} for a sketch burst.

For an overall burst time $t_b = t_g + t_p + t_d$, the constituent times can fluctuate (functional dependences determined from simulation). The theoretical maximum and minimum amplitudes were used to create a fit routine, allowing one to acquire from the mode amplitude $|E_j|$ the constituent times as a function of the parameters $k_j$, $\alpha_j$, and $\nu$. We find that these times are not deterministic, but instead are stochastic:

\begin{equation}
\begin{array}{l r l l}
\text{lag:} & t_l &= t_l^{(0)} &+ \, \delta t_l\\
\text{growth:} & t_g &= t_g^{(0)}\\
\text{plateau:} & t_p &= t_p^{(0)} &+ \, \delta t_p \\
\text{decay:} & t_d &= t_d^{(0)} &+ \, \delta t_d
\end{array} 
\end{equation}

where $\{\delta t_X\}$ denote stochastic terms. The fluctuation of each of the times has a well defined mean and standard deviation, and are analysed such that the mean is 0; here, we find that the lag, plateau, and decay times are highly stochastic. We find that at small $t_{\text{NL}}$ the mean plateau time $t_p^{(0)}$ is roughly constant, while the mean decay time $t_d^{(0)}$ is a linear function of $t_{\text{NL}}$, while the mean lag time $t_l^{(0)}$ is a non-linear function of $t_{\text{NL}}$.

\subsection{Linear phase}
When a single mode bursts, the electric field grows linearly via the bump-on-tail instability, provided that the mode lies on the positive slope of the `bump'. For $\nu = E_{\mathcal{S}} = E_{\mathcal{N}} = 0$, the frequency and growth rate in the linear phase are then determined by [\onlinecite{lifshitz81}]:

\begin{equation}
p + \dfrac{\alpha_j}{2} = \int\limits_{\Omega} \dfrac{v \, \partial_v F_0}{p + i k_j v} \dif  v
\end{equation}

where $\Omega$ is the suitable Landau contour for the problem, and $p \equiv \gamma_j - i \omega_j^{(0)}$, such that the overall linear growth rate for the $j^{\text{th}}$ mode is given by:

\begin{equation}
\gamma_j (k_j, \alpha_j) = \gamma_{j,L}(k_j, \alpha_j) - \dfrac{\alpha_j}{2}
\end{equation}

where the unperturbed linear growth rate $\gamma_{j,L}$ is equivalent to $\gamma_j$ in the absence of dissipation. 

\clearpage

\onecolumngrid

\begin{figure}
\centering
	\begin{subfigure}[t]{0.4\textwidth}
		\includegraphics[width=\textwidth]{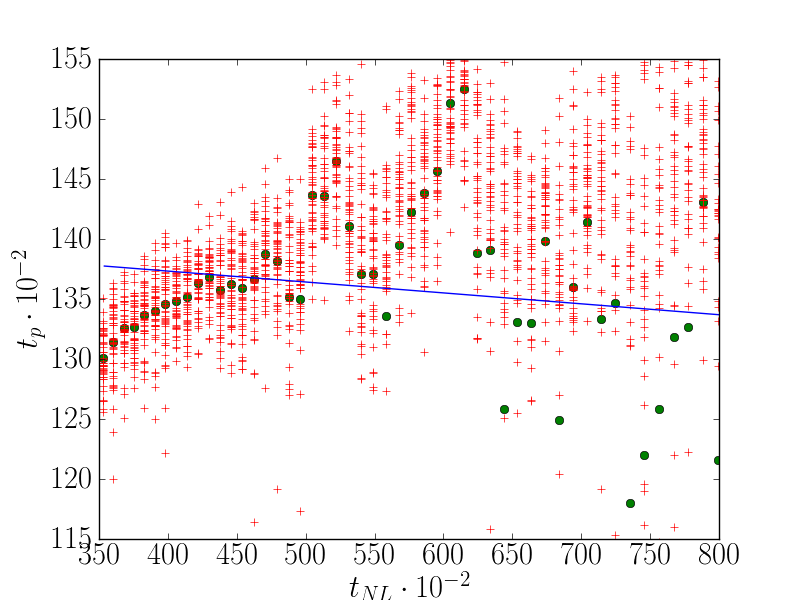}
		\subcaption{Plateau time $t_p$: approximately constant as a function of $t_{\text{NL}}$, with a relative stochastic fluctuation $\sigma_l / t_l ^{(0)} \sim 10^{-2}$.}
		\label{fig:tp}
	\end{subfigure}
	\hspace{40px}
	\begin{subfigure}[t]{0.4\textwidth}
		\includegraphics[width=\textwidth]{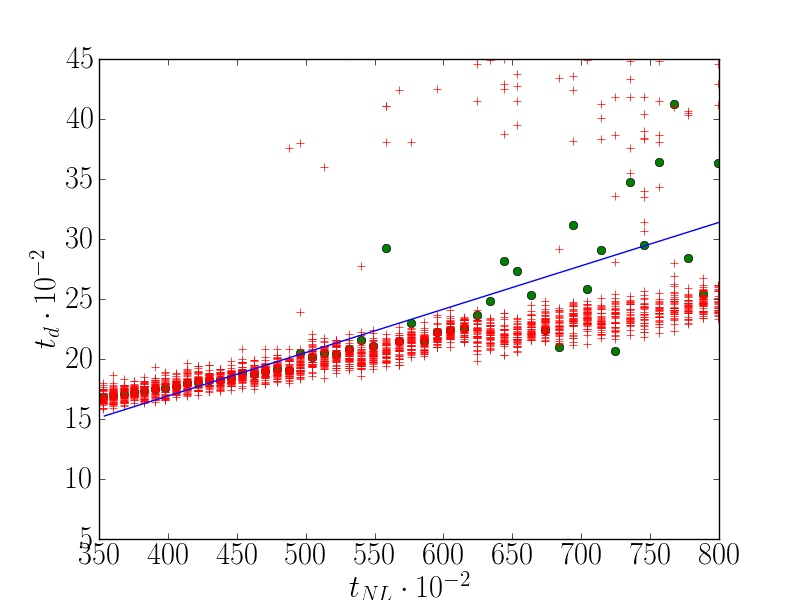}
		\subcaption{Decay time $t_p$: increases linearly as a function of $t_{\text{NL}}$, with a relative stochastic fluctuation $\sigma_l / t_d ^{(0)} \sim 10^{-1}$.} 
		\label{fig:td}
	\end{subfigure}
\caption{Graphs showing calculated values for $t_p(t_{\text{NL}})$ and $t_d(t_{\text{NL}})$ from simulations in \secref{sec:lifetime}. Observed values (`{\color{red} $+$}') and mean values (`{\color{green} $\bullet$}') from fit routines.}
\end{figure}

\twocolumngrid

The frequency $\omega_j^{(0)}$ is the initial frequency of the $j^{\text{th}}$ mode. If one solves \eqref{eq:ks4} for negligible current, on average:

\begin{equation}
|\delta E_j|_{\text{min}} \approx \dfrac{D_j}{\alpha_j / 2} \bigg[ 1 - e^{- \alpha_j \Delta t / 2} \bigg] + \mathcal{O}\left( e^{-(\alpha_j \Delta t)^2} \right)
\end{equation}

The simulated noise is static over a timestep, leading to an error which manifests as the term $\sim \exp(-\alpha_j \Delta t/2)$. 

One can interpret this physically as a finite bandwidth for the noise; we expect ITG turbulence spectral frequencies to typically be much slower than the plasma frequency [\onlinecite{citrin2017comparison}], however here we examine noise with a frequency spectrum that is typically below the plasma frequency, corresponding to high frequency turbulence. The peak value for the electric field is the non-linear saturation point, which can be approximated by the following value:[\onlinecite{berk1997spontaneous}]

\begin{equation}
|E_j|_{\text{max}} \approx ( \gamma_{j,L} )^2 
\end{equation}

Accordingly, as we expect exponential growth in the linear phase, the total time spent in the linear phase is given by:

\begin{equation}
(t_g^{(0)})_{\text{theory}} \approx \dfrac{2}{\gamma_j} \log \left[\dfrac{\alpha_j \gamma_{j,L}}{2 D_j [ 1 - \exp (- \alpha_j \Delta t / 2) ]}\right]
\end{equation}

Simulations were found to strongly agree with this value; we find $t_g^{(0)} = (-91.0 \pm 8.1) + (1.15 \pm 0.01) \cdot (t_g^{(0)})_{\text{theory}}$. The quantity $t_g$ does not appear to be stochastic; fluctuations in the value of $t_g$ as a function of PRNG seed are typically about 2 or 3 orders of magnitude lower than the mean value $t_g^{(0)}$. One finds that the accuracy improves at low $\alpha_j$; we find that this is in accordance with theory, as our value for $(t_g^{(0)})_{\text{theory}}$ assumes slow damping (small $\alpha_j \Delta t$).

At high $\alpha_j$, the expected growth time grows logarithmically until $\gamma_{j,L}$ dominates:

\[
\lim_{\alpha_j \to \infty} (t_g^{(0)})_{\text{theory}} = \dfrac{2}{\gamma_j} \left(\log \left[\dfrac{\gamma_{j,L}}{2 D_j} \right] + \log \alpha_j\right)
\]

Interestingly, one finds that even if the linear growth rate is non-zero, the seed electric field can prevent the mode from growing. This can be shown by setting the growth time to zero and solving for $\gamma_{j,L}$:

\[
(\gamma_j)_{\text{min}} \approx \dfrac{2 D_j}{\alpha_j} - \dfrac{\alpha_j}{2}
\]

If the linear growth rate is below this minimum, the seed electric field quenches the mode before it has a chance to burst; in order to preserve the true meaning of the linear growth rate, one should impose a limit on $D_j$ when using high $\alpha_j$:

\[
D_j \leq \dfrac{\alpha_j^2}{4}
\]

This hard limit on the noise level allows one to properly investigate simulations close to the linear stability boundary.

\clearpage

\onecolumngrid

\begin{figure}
\centering
	\begin{subfigure}[t]{0.3\textwidth}
		\begin{subfigure}[t]{\textwidth}
			\includegraphics[width=\textwidth]{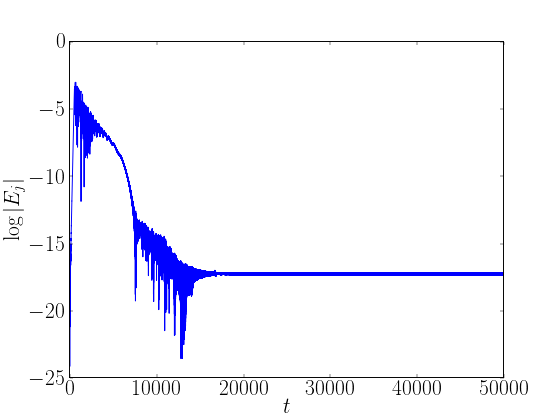}
			\label{fig:nu6R00}
		\end{subfigure}
		\begin{subfigure}[t]{\textwidth}
			\includegraphics[width=\textwidth]{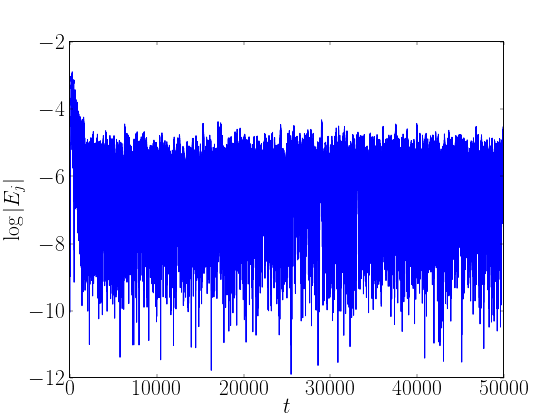}
			\label{fig:nu5R20}
		\end{subfigure}
		\subcaption{Low effective collisionality: the mode amplitude $|E_j|$ undergoes a single bursting event at $t \approx 0$, corresponding to mode chirping. Top plot shows $\nu = 10^{-6}$, $R_j = 0$; bottom plot shows $\nu = 10^{-5}$, $R_j = 10^{-2}$.}
		\label{lnElow}
	\end{subfigure}
	\hspace{10px}
	\begin{subfigure}[t]{0.3\textwidth}
		\begin{subfigure}[t]{\textwidth}
			\includegraphics[width=\textwidth]{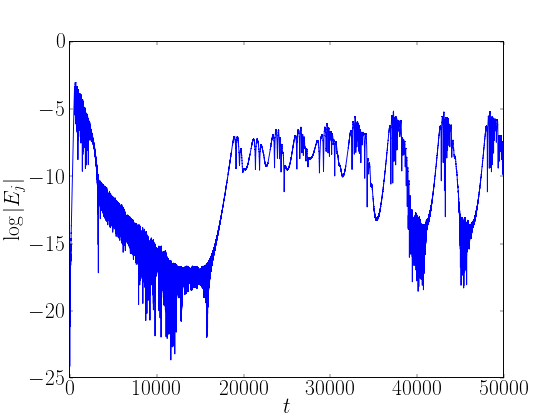}
			\label{fig:nu5R00}
		\end{subfigure}
		\begin{subfigure}[t]{\textwidth}
			\includegraphics[width=\textwidth]{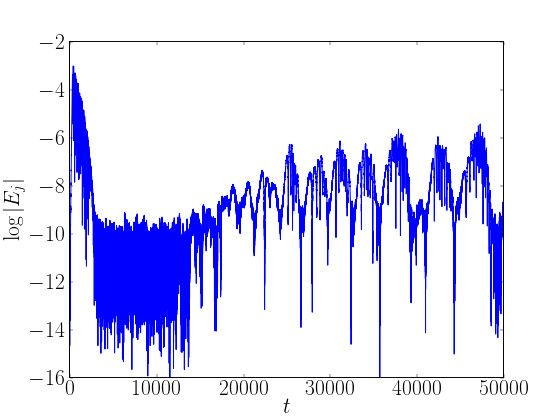}
			\label{fig:nu5R40}
		\end{subfigure}
		\subcaption{Medium effective collisionality: $|E_j|$ undergoes repeated bursting events ($t \approx \{0, 22000, 24500, \dots\}$), each corresponding to mode chirping. Top plot shows $\nu = 10^{-5}$, $R_j = 0$; bottom plot shows $\nu = 10^{-5}$, $R_j=10^{-4}$.}
		\label{lnEmed}
	\end{subfigure}
	\hspace{10px}
	\begin{subfigure}[t]{0.3\textwidth}
		\begin{subfigure}[t]{\textwidth}
			\includegraphics[width=\textwidth]{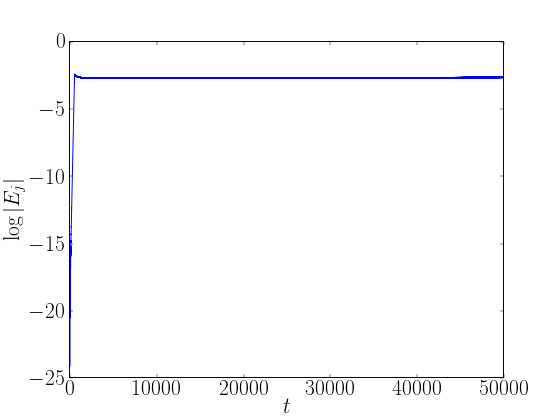}
			\label{fig:nu2R00}
		\end{subfigure}
		\begin{subfigure}[t]{\textwidth}
			\includegraphics[width=\textwidth]{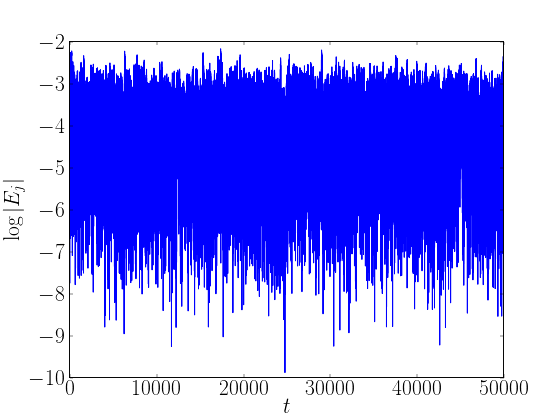}
			\label{fig:nu5R10}
		\end{subfigure}
		\subcaption{High effective collisionality: the system does not undergo mode chirping. Top plot shows $\nu = 10^{-2}$, $R_j = 0$; bottom plot shows $\nu = 10^{-5}$, $R_j=10^{-1}$.}
		\label{lnEhigh}
	\end{subfigure}
\caption{Plots of $\log |E_j|(t)$ from linearly unstable simulations in \secref{sec:suppression}. Asymptotic behaviours for a single mode are shown for deterministic cases with varying collisionality $\nu$ (top plots), and stochastic cases with fixed collisionality $\nu = 10^{-5}$ (bottom plots).}
\label{fig:lnE}
\end{figure}

\twocolumngrid 

\subsection{Non-linear phase}
Once the mode reaches the non-linear saturation point, resonance broadening occurs, flattening the background distribution function $f_{\text{ion},0}$ in the close vicinity of the resonant phase velocity $\omega_j^{(0)}/k_j$. If the mode is marginally unstable, to the extent that: [\onlinecite{berk1997spontaneous}]

\begin{equation}
\label{eq:bbp}
\dfrac{\alpha_j}{2} > 0.2 \gamma_{j,L}
\end{equation}

a phase-space bifurcation in the form of a hole and clump form on $f_{\text{ion},0}$.


For diffusive Fokker-Planck collisions, the time spent in the plateau and decay regions has previously been shown to be a function of the timescale[\onlinecite{berk1997spontaneous}] $t_{\text{NL}} = (\gamma_{j,L})^2/[(\omega_j^{(0)})^2 \nu]$. \\

We take first order Taylor expansions in $t_{\text{NL}}$ as follows:

\begin{equation}
t_X^{(0)} = a_X + b_X t_{\text{NL}} + \mathcal{O} (t_{\text{NL}}^2) = \dfrac{1}{50} \sum\limits_{\text{seed}} t_X
\end{equation}

where we sum over 50 PRNG seeds. From the simulations, $t_l^{(0)}$ appears to be a non-linear function of $t_{\text{NL}}$ while $t_d^{(0)}$ appears to be linear.

The mean plateau time $t_p^{(0)}$ appears to be constant at low $t_{\text{NL}}$; however at high $t_{\text{NL}}$, large error in the linear fit reduces our ability to determine the mean time.

We find that $a_d = (2.56 \pm 2.01) \times 10^2$, $b_d = (3.62 \pm 0.35) \times 10^{-2})$, and $a_p = (1.41 \pm 0.04) \times 10^4$. One should note that the errors here are errors in the linear fit to mean values; they represent confidence in the functional dependence on $t_{\text{NL}}$, not the stochasticity. We find that $\mathcal{O}(b_p) = 10^{-3}$, allowing us to state $t_p^{(0)} \approx a_p$ for $t_{\text{NL}} << 10^6$.

\clearpage

\onecolumngrid

\begin{figure}
\centering
	\begin{subfigure}[t]{0.3\textwidth}
		\begin{subfigure}[t]{\textwidth}
			\includegraphics[width=\textwidth]{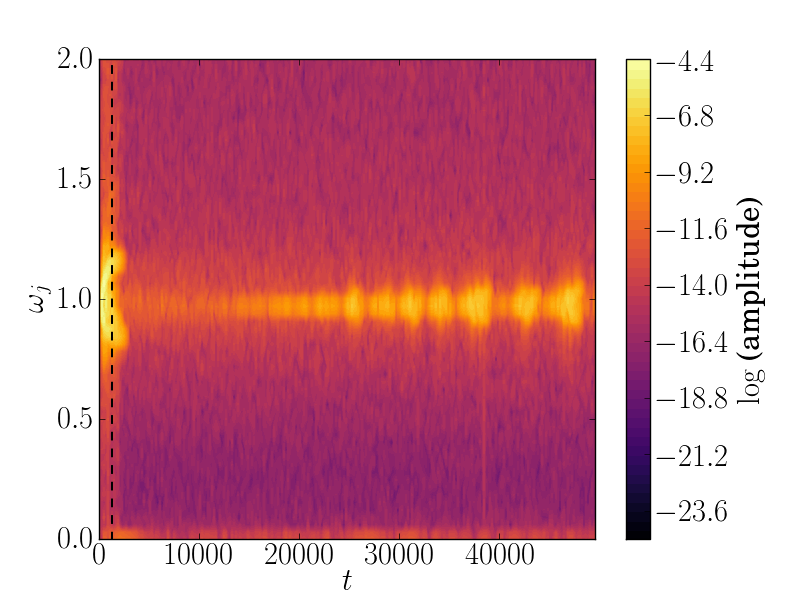}
			\label{fig:wR40}
		\end{subfigure} \\
		\begin{subfigure}[t]{\textwidth}
			\includegraphics[width=\textwidth]{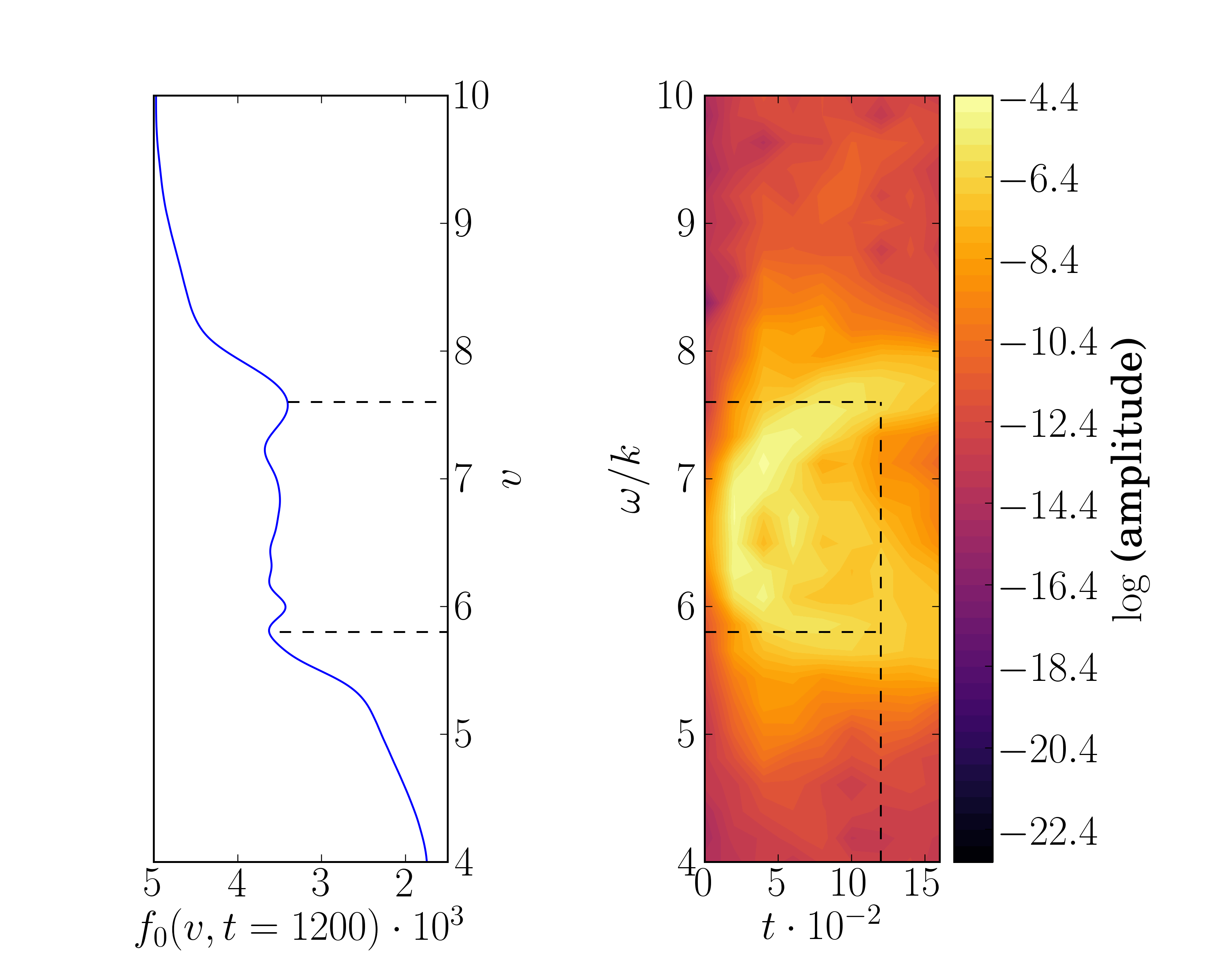}
			\label{fig:fR40}
		\end{subfigure}
		\subcaption{$R_j = 10^{-4}$: noise in $f_{\text{tur}}$ produces an electric field, but repeated bursting still occurs.}
		\label{fig:wfR40}
	\end{subfigure}
	\hspace{10px}
	\begin{subfigure}[t]{0.3\textwidth}
		\begin{subfigure}[t]{\textwidth}
			\includegraphics[width=\textwidth]{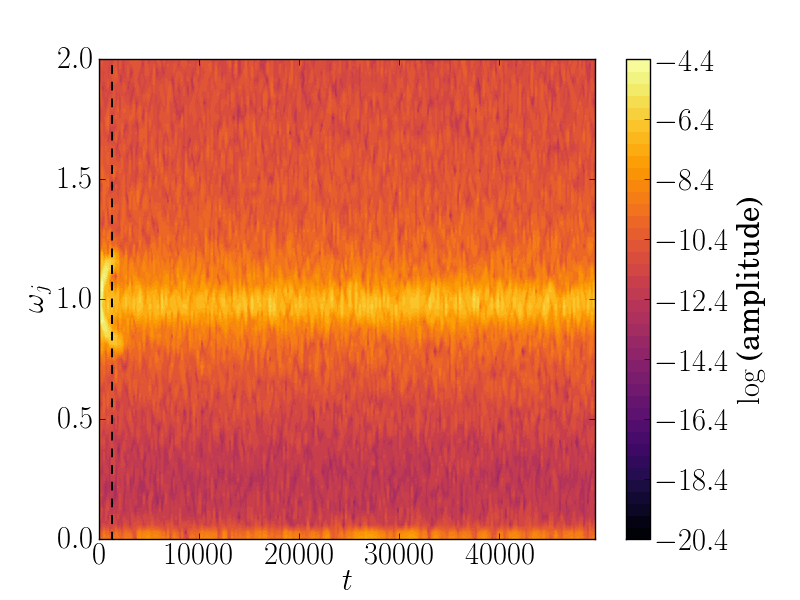}
			\label{fig:wR20}
		\end{subfigure} \\
		\begin{subfigure}[t]{\textwidth}
			\includegraphics[width=\textwidth]{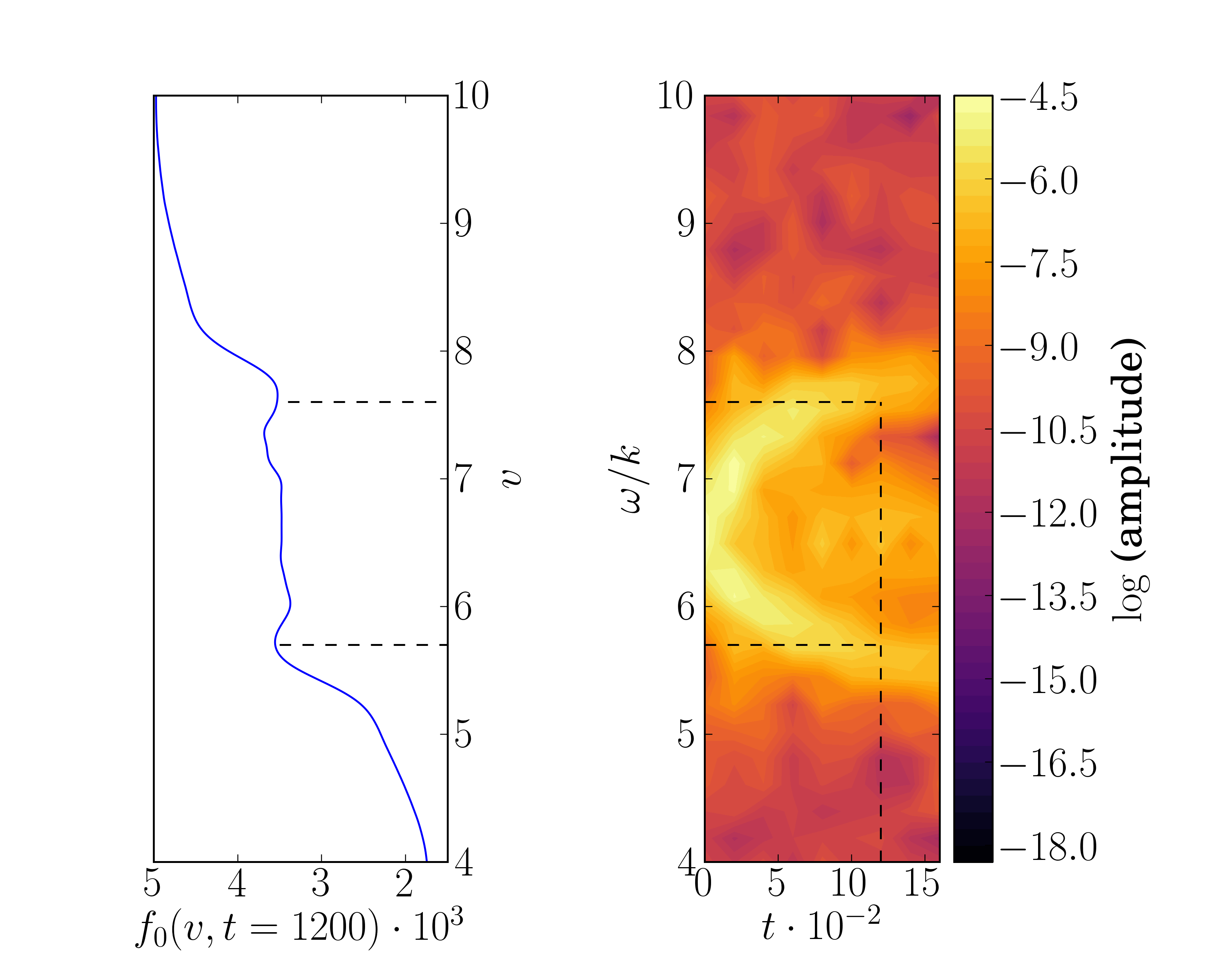}
			\label{fig:fR20}
		\end{subfigure}
		\subcaption{$R_j = 10^{-2}$: noise in $f_{\text{tur}}$ produces an electric field which prevents repeated bursts from occuring, but does not prevent the mode from initially chirping.}
		\label{fig:wfR20}
	\end{subfigure}
	\hspace{10px}
	\begin{subfigure}[t]{0.3\textwidth}
		\begin{subfigure}[t]{\textwidth}
			\includegraphics[width=\textwidth]{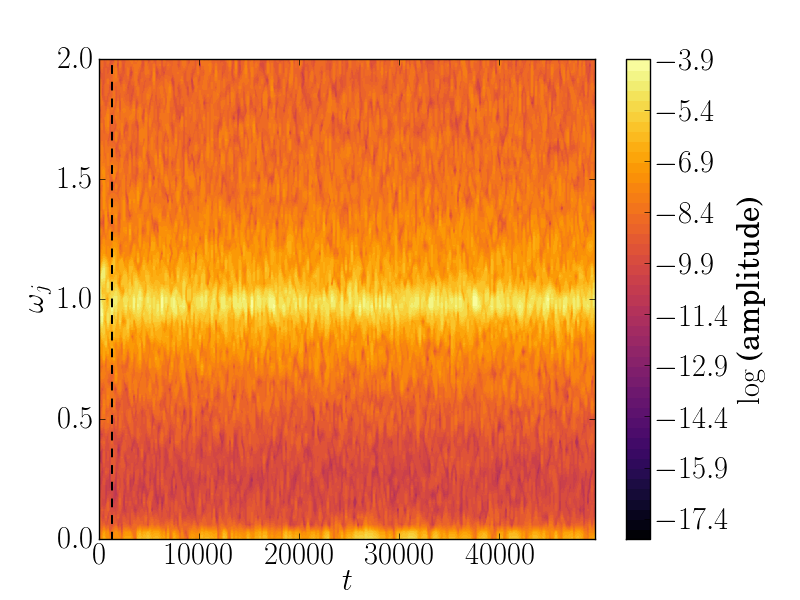}
			\label{fig:wR10}
		\end{subfigure} \\
		\begin{subfigure}[t]{\textwidth}
			\includegraphics[width=\textwidth]{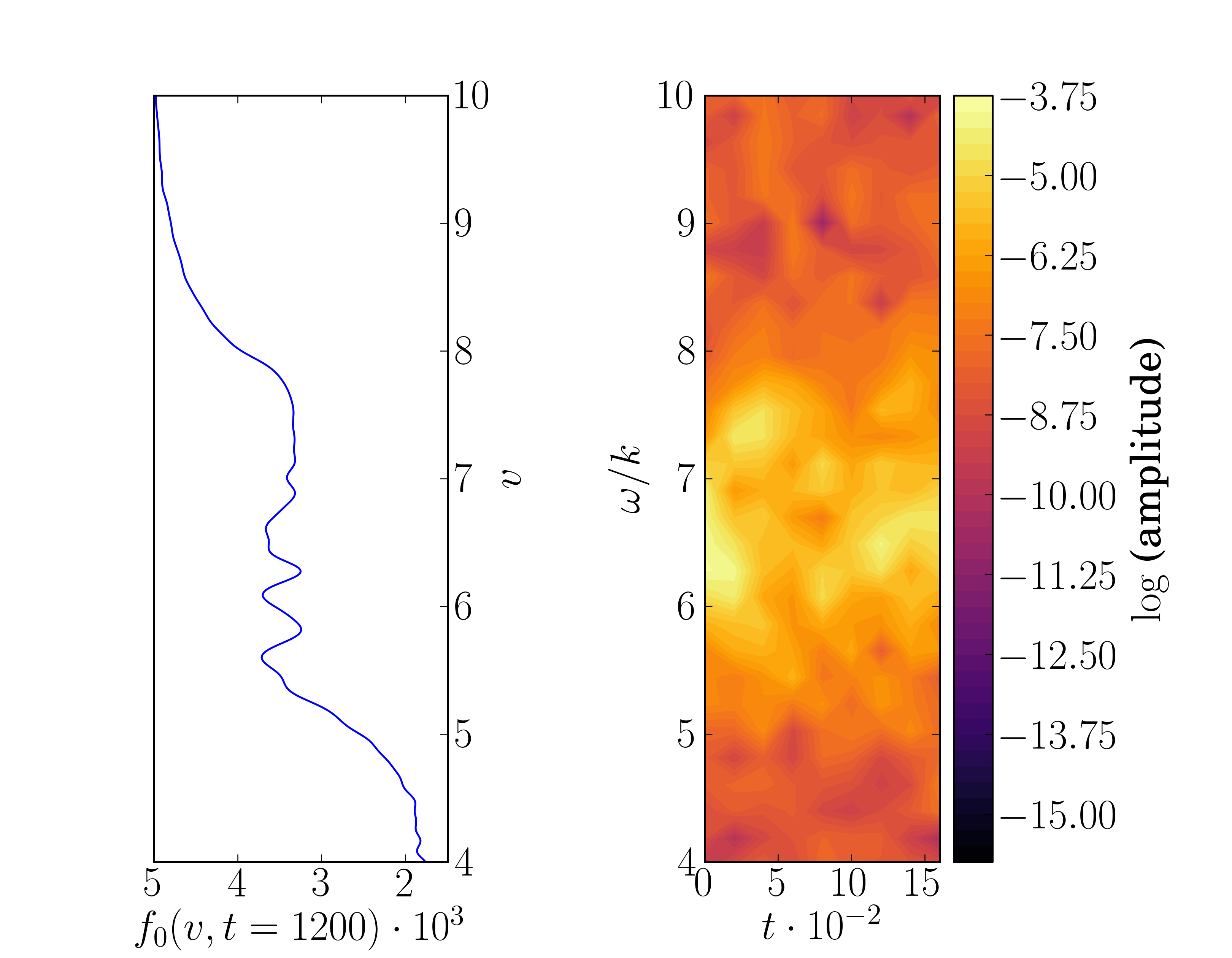}
			\label{fig:fR10}
		\end{subfigure}
		\subcaption{$R_j = 10^{-1}$: noise in $f_{\text{tur}}$ produces an electric field which prevents a hole and clump from forming; the system is non-linearly stable.}
		\label{fig:wfR10}
	\end{subfigure}
\caption{Plots of $\omega_j(t)$ and $f_0(v) = \bar{f}_{\text{ion}}(x,v,t=1200)$ (top plots and bottom plots respectively) from simulations in \secref{sec:suppression} with $\nu = 10^{-5}$; the relative stochasticity $R_j$ is varied between simulations. Black dotted lines in the bottom plots show the existence (or lack of) hole and clump at $t = 1200$.}
\label{fig:sup}
\end{figure}

\twocolumngrid

\subsection{Burst stochasticity}
We once again take first order Taylor expansions in $t_{\text{NL}}$, but now examining the standard deviations in $\{t_X\}$:

\begin{equation}
\sigma_X = c_X + d_X t_{\text{NL}} + \mathcal{O} (t_{\text{NL}}^2) = \dfrac{1}{50} \sum\limits_{\text{seed}} \delta t_X^2
\end{equation}

where we once again sum over 50 PRNG seeds. We find that generally, $\sigma_X$ does not appear to be a function of $t_{\text{NL}}$. The lag time $t_l$ is strongly stochastic, with $\sigma_l / t_l ^{(0)} \sim 10^0$. This is in accordance with theory, as at very low amplitude, $\partial_t|E_j|$ is strongly dependent on the noise term, which is stochastic. 

The plateau exists while the hole and clump have a static population of particles, and therefore, once the phase-space structures dissipate, the mode drops significantly in amplitude. The time spent in this region, $t_p$, is stochastic; as is shown in \figref{fig:tp}, the relative fluctuation $\sigma_p / t_p^{(0)} \sim 10^{-2}$. We conclude that the point at which this occurs is stochastic, leading to a stochastic life time of the hole and clump.

The growth and decay times $t_g$ and $t_d$ are defined by the minimum and maximum mode amplitude. Therefore, any stochastic behaviour reflects fluctuation in the growth rate and decay rate of the mode. We find that $\sigma_g / t_g$ is negligible, implying that mode growth is not stochastic, as one might expect. However, we find that $\sigma_d / t_d^{(0)} \sim 10^{-1}$, implying a large fluctuation in the decay rate of the mode (see \figref{fig:td}). 

\section{\label{sec:suppression} Stochastic suppression of hole and clump}
Here, we consider a case with a wholly non-stochastic electric field ($\epsilon_E = 0$), and examined the effect of a stochastic distribution function $f_{\text{tur}}$ on mode chirping.

One finds that in cases of high collisionality, we must enforce $\nu \lessapprox \Delta v^2/\Delta t$ to avoid numerical inaccuracies where collisions dissipate structures much faster than the timestep. We fix the linear growth rate to $\gamma_j = \alpha_j/2 = 0.0534$, to reduce the parameter space while still allowing hole and clump formation (see \eqref{eq:bbp}). We fix $D$ at $10^{-7}$ and $1/(1 + \epsilon_{f,j}) = 98\%$ to investigate a small electric field drive and a small turbulent population. We define the relative stochasticity as $R_j \equiv \log_{10} (\sigma_{f,j}/\sigma^{(\text{max})}_{f,j})$.

In theory, $f_{\text{tur}}$ can affect $\delta E$ via \eqref{eq:ks4}. As $(W \cdot f_{\text{tur}}) \in \mathbb{I}$, one can see that real stochastic noise will produce an imaginary stochastic term in \eqref{eq:ks4}, which will lead to a scrambling of the phase of coherent structures with wavenumber $k_j$. Consequently, the coherence of hole and clump structures can be destroyed, allowing for shear dissipation via dispersion. 

In contrast with $E_{\mathcal{N}}$, as $f_{\text{tur}}$ exchanges no energy on pseudorandom time average with $E$, it can instantaneously create perturbations in $E$ which cannot be induced by finite $E_{\mathcal{N}}$.

First, we show 3 simulations with varying collisionality $\log_{10} \nu = \{-2, -5, -6\}$, and no noise ($R_j = 0$). A timestep of $\Delta t = 0.01$ to allow us safe exploration of $\nu \sim 10^{-2}$. As is shown in \figref{fig:lnE}, high collisionality ($\nu = 10^{-2}$) suppresses mode chirping, medium collisionality ($\nu = 10^{-5}$) allows for repeated bursting, and low collisionality ($\nu = 10^{-6}$) allows for only a single event.

Next, we highlight the effect of $\sigma_{f,j}$ on the asymptotic behaviour of the mode in \figref{fig:lnE} and \figref{fig:sup}; we show results from 3 simulations with $\nu = 10^{-5}$ using a coarser timestep of $\Delta t = 0.1$, and $R_j = \{-1,-2,-4\}$. For low stochasticity ($R_j = 10^{-4}$), the effective collisionality increases; we still observe repeated bursting, however the period between repeated bursts is characteristic of simulations with $\nu \sim 10^{-4}$. Here, $f_{\text{tur}}$ affects the stability of $f_{\text{ion}}$, but does not make the fast-ion population stable; as is shown in \figref{fig:wfR40}, the initial hole and clump is undisturbed, and repeated bursts still occur.

For medium stochasticity ($R_j = 10^{-2}$), we find that the repeated bursts are suppressed. We give two equivalent explanations: the electric field produced by $f_{\text{tur}}$ approaches the maximum amplitude of the repeated bursts, saturating them. As is shown in \figref{fig:wfR20}, the initial hole and clump still exists, but repeated bursts do not occur.

As we increase to high stochasticity ($R_j = 10^{-1}$), we find that the initial burst is suppressed. We find that this is when the electric field produced by $f_{\text{tur}}$ has an amplitude close to the non-linear saturation point; at this point, the mode does not resonate, even though it is unstable. Alternatively, $f_{\text{tur}}$ prevents the mode from bursting by increasing the effective collisionality; as is shown in \figref{fig:wfR10}, the electric field produced by $f_{\text{tur}}$ creates features on the spatially averaged distribution $f_0(v) = \bar{f}_{\text{ion}}(x,v,t=1200)$ which affect the hole and clump.

\section{Conclusions and future work}
In the case of weak stochasticity, we conclude that mode chirping is not wholly deterministic; the shape of the burst in mode amplitude can be determined on average analytically, but varies depending on the noise seed employed. We hope that this theory will lead to further analytical work, allowing one to freely predict the shape of a burst as a function of the whole parameter space: it is reasonable to assume that the empirical coefficients $\{a_X\}$ and $\{b_X\}$ are dependent on the shape of $F_0$.

Further work expanding this theory to include the time between bursts could lead to predictions of the burst frequency between Alfv\'{e}n mode chirping events in tokamaks, which would allow for a greater understanding of fast ion loss. We plan to couple predictions about the shape of a given burst to analytic theory by Berk, Breizman and Petviashvili [\onlinecite{berk1997spontaneous}] to create an analytical theory of hole-and-clump destabilisation in future work.

We conclude that the lifetime of a hole and clump is stochastic, and that the decay rate of the mode is also stochastic. Again, we hope that the accuracy of analysis for $t_p$ and $t_d$ improves with further work; in reality, the plateau region has a slight negative slope. It is our belief that an upgraded model with a negative gradient for $t_p$ would yield greater accuracy on the non-linear dependence of $t_p$ on $t_{\text{NL}}$.

We find that increasing the stochasticity in the system is initially equivalent to increasing the effective diffusive collisionality. This is as one might expect from theory; stochasticity in the turbulent distribution function or electric field affects the damping term $\alpha E$, which is analogous to the energetic particle drive $\dif f/\dif p_{\varphi}$. Accordingly, simulations with increased stochasticity produce similar overall results to theory based on stochastic perturbations to momentum scattering via microturbulence-induced radial diffusion.[\onlinecite{lang2011nonlinear}] However, an important nuance appears when considering repeated bursting; low amplitude repeated bursts are saturated by the electric field produced by $f_{\text{tur}}$, leading to an asymptotic behaviour characteristic of a decrease in the effective collisionality. 

As a result, we posit that in a given regime, an increase in micro-turbulence leads to an anomalous decrease in the effective collisionality; it is implied that in this regime, microturbulence reduces the ability for the distribution function to reconstitute via pitch-angle scattering.

\section{Acknowledgments}
One of the authors (BJQW) was funded by the EPSRC Centre for Doctoral Training in Science and Technology of Fusion Energy grant EP/L01663X. This work made use of the `York Advanced Research Computing Cluster (YARCC)' at The University of York. This work made use of the facilities of N8 HPC Centre of Excellence, provided and funded by the N8 consortium and EPSRC (grant no. EP/K000225). RV would like to thank B. N. Breizman for useful discussions.

\small
\bibliographystyle{unsrt}
\bibliography{biblio}

\begin{thebibliography}{10}

\bibitem{lang2011nonlinear}
J.~Lang and G-Y. Fu.
\newblock {Nonlinear simulation of toroidal Alfv{\'e}n eigenmode with
  microturbulence-induced radial diffusion}.
\newblock {\em Physics of Plasmas}, 18(5):055902, 2011.

\bibitem{heidbrink2008basic}
W.~W. Heidbrink.
\newblock {Basic physics of Alfv{\'e}n instabilities driven by energetic
  particles in toroidally confined plasmas}.
\newblock {\em Physics of Plasmas}, 15(5):055501, 2008.

\bibitem{pinches2004role}
S.~D. Pinches et~al.
\newblock {The role of energetic particles in fusion plasmas}.
\newblock {\em Plasma physics and controlled fusion}, 46(12B):B187, 2004.

\bibitem{fredrickson2006fast}
E.~D. Fredrickson, N.~N. Gorelenkov, R.~E. Bell, J.~E. Menard, A.~L. Roquemore,
  S.~Kubota, N.~A. Crocker, and W.~Peebles.
\newblock {Fast ion loss in a ‘sea-of-TAE’}.
\newblock {\em Nuclear fusion}, 46(10):S926, 2006.

\bibitem{gorelenkov2014energetic}
N.~N. Gorelenkov, S.~D. Pinches, and K.~Toi.
\newblock {Energetic particle physics in fusion research in preparation for
  burning plasma experiments}.
\newblock {\em Nuclear Fusion}, 54(12):125001, 2014.

\bibitem{berk1997spontaneous}
H.~L. Berk, B.~N. Breizman, and N.~V. Petviashvili.
\newblock {Spontaneous hole-clump pair creation in weakly unstable plasmas}.
\newblock {\em Physics Letters A}, 234(3):213--218, 1997.

\bibitem{breizman1997critical}
B.~N. Breizman, H.~L. Berk, M.~S. Pekker, F.~Porcelli, G.~V. Stupakov, and
  K.~L. Wong.
\newblock {Critical nonlinear phenomena for kinetic instabilities near
  threshold}.
\newblock {\em Physics of Plasmas}, 4(5):1559--1568, 1997.

\bibitem{wang2013hole}
H.~Wang, Y.~Todo, C.~C. Kim, et~al.
\newblock {Hole-clump pair creation in the evolution of
  energetic-particle-driven geodesic acoustic modes}.
\newblock {\em Physical review letters}, 110(15):155006, 2013.

\bibitem{duarte2017prediction}
V.~N. Duarte, H.~L. Berk, N.~N. Gorelenkov, W.~W. Heidbrink, G.~J. Kramer,
  R.~Nazikian, D.~C. Pace, M.~Podesta, B.~J. Tobias, and M.~A. Van~Zeeland.
\newblock {Prediction of nonlinear evolution character of
  energetic-particle-driven instabilities}.
\newblock {\em Nuclear Fusion}, 57(5):054001, 2017.

\bibitem{duarte17-2}
V.~N. Duarte, H.~L. Berk, N.~N. Gorelenkov, W.~W. Heidbrink, G.~J. Kramer,
  R.~Nazikian, D.~C. Pace, M.~Podesta, B.~J. Tobias, and M.~A. Van~Zeeland.
\newblock {Theory and observation of the onset of nonlinear structures due to
  eigenmode destabilization by fast ions in tokamaks}.
\newblock {\em Phys. Plas.}, 24(000000), 2017.

\bibitem{vanzeeland17}
M.~Van~Zeeland et~al.
\newblock {Alfv{\'e}n Eigenmodes and Fast Ion Transport in Negative
  Triangularity DIII-D Plasmas}.
\newblock 2017.

\bibitem{breizman2010nonlinear}
B.~N. Breizman.
\newblock {Nonlinear travelling waves in energetic particle phase space}.
\newblock {\em Nuclear Fusion}, 50(8):084014, 2010.

\bibitem{berk1995numerical}
H.~L. Berk, B.~N. Breizman, and M.~Pekker.
\newblock {Numerical simulation of bump-on-tail instability with source and
  sink}.
\newblock {\em Physics of Plasmas}, 2(8):3007--3016, 1995.

\bibitem{vann2003fully}
R.~G.~L. Vann, R.~O. Dendy, G.~Rowlands, T.~D. Arber, and N.~d'Ambrumenil.
\newblock {Fully nonlinear phenomenology of the Berk--Breizman augmentation of
  the Vlasov--Maxwell system}.
\newblock {\em Physics of Plasmas}, 10(3):623--630, 2003.

\bibitem{degol10}
A.~P. De-Gol.
\newblock {\em {Nonlinear wave-particle phenomena in a Berk-Breizman
  Vlasov-Maxwell system}}.
\newblock PhD thesis, University of York, 2010.

\bibitem{lang2010nonlinear}
J.~Lang, G-Y. Fu, and Y.~Chen.
\newblock {Nonlinear simulation of toroidal Alfv{\'e}n eigenmode with source
  and sink}.
\newblock {\em Physics of Plasmas}, 17(4):042309, 2010.

\bibitem{ghantous2014comparing}
K.~Ghantous, H.~L. Berk, and N.~N. Gorelenkov.
\newblock {Comparing the line broadened quasilinear model to Vlasov code}.
\newblock {\em Physics of Plasmas}, 21(3):032119, 2014.

\bibitem{lilley2009destabilizing}
M.~K. Lilley, B.~N. Breizman, and S.~E. Sharapov.
\newblock {Destabilizing effect of dynamical friction on fast-particle-driven
  waves in a near-threshold nonlinear regime}.
\newblock {\em Physical review letters}, 102(19):195003, 2009.

\bibitem{linz1997nonlinear}
S.~J. Linz.
\newblock {Nonlinear dynamical models and jerky motion}.
\newblock {\em American Journal of Physics}, 65(6):523--526, 1997.

\bibitem{poincare1885courbes}
H.~Poincar{\'e}.
\newblock {Sur les courbes d{\'e}finies par les {\'e}quations
  diff{\'e}rentielles (III)}.
\newblock {\em Journal de math{\'e}matiques pures et appliqu{\'e}es},
  1:167--244, 1885.

\bibitem{bendixson1901courbes}
I.~Bendixson.
\newblock {Sur les courbes d{\'e}finies par des {\'e}quations
  diff{\'e}rentielles}.
\newblock {\em Acta Mathematica}, 24(1):1--88, 1901.

\bibitem{chlouverakis2006chaotic}
K.~E. Chlouverakis and J.~C. Sprott.
\newblock {Chaotic hyperjerk systems}.
\newblock {\em Chaos, Solitons \& Fractals}, 28(3):739--746, 2006.

\bibitem{arber2002critical}
T.~D. Arber and R.~G.~L. Vann.
\newblock {A critical comparison of Eulerian-grid-based Vlasov solvers}.
\newblock {\em Journal of computational physics}, 180(1):339--357, 2002.

\bibitem{fftw}
M.~Frigo and S.~G. Johnson.
\newblock {The Design and Implementation of FFTW3}.
\newblock {\em Proc. IEEE}, 93(2):216--231, 2005.

\bibitem{lesur2012nonlinear}
M.~Lesur and Y.~Idomura.
\newblock {Nonlinear categorization of the energetic-beam-driven instability
  with drag and diffusion}.
\newblock {\em Nuclear Fusion}, 52(9):094004, 2012.

\bibitem{lesur2016nonlinear}
M.~Lesur et~al.
\newblock {Nonlinear excitation of subcritical fast ion-driven modes}.
\newblock {\em Nuclear Fusion}, 56(5):056009, 2016.

\bibitem{courant1928partiellen}
R.~Courant, K.~Friedrichs, and H.~Lewy.
\newblock {{\"U}ber die partiellen Differenzengleichungen der mathematischen
  Physik}.
\newblock {\em Mathematische annalen}, 100(1):32--74, 1928.

\bibitem{colella1984piecewise}
P.~Colella and P.~R. Woodward.
\newblock {The piecewise parabolic method (PPM) for gas-dynamical simulations}.
\newblock {\em Journal of computational physics}, 54(1):174--201, 1984.

\bibitem{lifshitz81}
E.~M. Lifshitz and L.~P. Pitaevskii.
\newblock {\em {Physical Kinetics}}.
\newblock Pergamon, 1st edition, 1981.

\bibitem{citrin2017comparison}
J.~Citrin, H.~Arnichand, J.~Bernardo, C.~Bourdelle, X.~Garbet, F.~Jenko,
  S.~Hacquin, M.~J. Pueschel, and R.~Sabot.
\newblock {Comparison between measured and predicted turbulence frequency
  spectra in ITG and TEM regimes}.
\newblock {\em Plasma Physics and Controlled Fusion}, 59(6), 2017.

\bibitem{burgess}
M.~Burgess.
\newblock {\em {Classical Covariant Fields}}.
\newblock Cambridge University Press, 1st edition, 2002.

\bibitem{schur1889neue}
F.~Schur.
\newblock {Neue Begr{\"u}ndung der Theorie der endlichen
  Transformationsgruppen}.
\newblock {\em Mathematische Annalen}, 35(1-2):161--197, 1889.

\bibitem{strang1968construction}
G.~Strang.
\newblock {On the construction and comparison of difference schemes}.
\newblock {\em SIAM Journal on Numerical Analysis}, 5(3):506--517, 1968.

\bibitem{cheng1976integration}
C-Z. Cheng and G.~Knorr.
\newblock {The integration of the Vlasov equation in configuration space}.
\newblock {\em Journal of Computational Physics}, 22(3):330--351, 1976.

\end{thebibliography}

\begin{appendix}
\section{\label{app:bbm} Resonant damping}
The classical Lagrangian and Hamiltonian densities of the electromagnetic field are given by [\onlinecite{burgess}]:

\begin{equation}
\begin{array}{r c l}
\mathcal{L} = - \dfrac{F^{\alpha \beta} F_{\alpha \beta}}{4 \mu_0} - A_{\alpha} J^{\alpha} &;& \mathcal{H}_0 = \Pi^{\beta \alpha} \partial_{\beta} A_{\alpha} - \mathcal{L}_0
\end{array}
\end{equation}

where $F^{\alpha \beta}$ is the electromagnetic force tensor, $\mu_0$ is the permeability of free space, $J^{\alpha}$ is the four-current. $A_\alpha$ and $\Pi^{\beta \alpha}$ are the four-potential and conjugate $\Pi$-tensor:

\begin{equation}
\begin{array}{r c l}
A_{\alpha} = (\phi/c, \mathbf{A}) &;& \Pi^{\beta \alpha} = \dfrac{\partial \mathcal{L}}{\partial (\partial_{\beta} A_{\alpha})}
\end{array}
\end{equation}

where $\phi$ is the electric scalar potential, and $\mathbf{A}$ is the magnetic vector potential. 

\subsection{Augmentation tensor, $G^{\beta \alpha}$}
Let us define $\mathcal{L}(A_{\alpha}, \partial_{\beta} A_{\alpha}) = \mathcal{L}_0 + \delta \mathcal{L}$. Then, $\mathcal{H}(A_{\alpha}, \Pi^{\beta \alpha})$ is given by the appropriate Legendre transformation:

\[
\renewcommand{\arraystretch}{2.5}
\begin{array}{r l}
\mathcal{H} &= \Pi^{\beta \alpha} \partial_{\beta} A_{\alpha} - \mathcal{L} \\
&= \underbrace{\left[ \dfrac{\partial \mathcal{L}_0}{\partial(\partial_{\beta} A_{\alpha})} \partial_{\beta} A_{\alpha} - \mathcal{L}_0 \right]}_{\mathcal{H}_0} + \underbrace{\left[ \dfrac{\partial (\delta \mathcal{L})}{\partial(\partial_{\beta} A_{\alpha})} \partial_{\beta} A_{\alpha} - \delta \mathcal{L} \right]}_{\delta \mathcal{H}}
\end{array}
\]

where we define $\mathcal{H}_0: \delta L = 0$. We seek the perturbation $\delta \mathcal{H} = 0$ so as to preserve the canonical form of the Hamiltonian. Therefore:

\[
\delta \mathcal{L} =  \dfrac{\partial (\delta \mathcal{L})}{\partial(\partial_{\beta} A_{\alpha})} \partial_{\beta} A_{\alpha}
\]

This trivial partial differential equation solves to give:

\[
\delta \mathcal{L} = G^{\beta \alpha} (A_{\mu}) \cdot \partial_{\beta} A_{\alpha}
\]

where the augmentation tensor $G^{\beta \alpha} (A_{\mu})$ preserves the Lorentz invariance of the Lagrangian density, but is only a function of the four-potential.

\subsection{Canonical form}
\label{app:canon}
From $\mathcal{L} = \mathcal{L} _0+ \delta \mathcal{L}$, the augmented Maxwell's equations for the system are given by the generalized Euler-Lagrange equations:

\begin{equation}
\partial_{\beta} \left[ \dfrac{\partial \mathcal{L}}{\partial(\partial_\beta A_{\alpha})}\right] - \dfrac{\partial \mathcal{L}}{\partial A_{\alpha}} = 0
\end{equation}

By examining a 1D Cartesian space with no $\mathbf{B}$-field, we seek an augmentation tensor that satisfies:

\begin{equation}
- \alpha E = \dfrac{1}{\epsilon_0} \left[ A_\mu \dfrac{\partial G^{0 \mu}}{\partial A_x} + \partial_x A_\mu \dfrac{\partial G^{1 \mu}}{\partial A_x} - (G^{01} + G^{11}) \right]
\end{equation}

Doing so allows the augmentation to the Gauss-Ampere law \emph{a posteriori} to manifest as a Berk-Breizman sink of energy via a global dissipation channel. [\onlinecite{berk1997spontaneous,vann2003fully}] This is non trivially satisfied, but the simplest case is when $\partial G^{\nu \mu} / \partial A = 0$, and:

\begin{equation}
G^{01} + G^{11} = - \alpha \epsilon_0 (\partial_x \phi + \dot{A})
\end{equation}

One can also show that further constraints on the augmentation tensor allow Gauss' law to retain the exact same form as before. Therefore, this family of augmentation tensors produce the modified Maxwell-Ampere law and preserve the canonical form of the energy density:

\begin{equation}
\begin{array}{r c l}
\partial_t E = - \dfrac{J}{\epsilon_0} - \alpha E &;& U = \dfrac{1}{\epsilon_0} \displaystyle\int\limits_{-\infty}^{\infty} E^2 \dif x
\end{array}
\end{equation}

The Lorentz force on charged particles due to the electromagnetic field is given by the particle Lagrangian.[\onlinecite{burgess}] If we examine a single particle:

\begin{equation}
L_p = \left[\dfrac{1}{2} m u_{\mu} u^\mu + q u^\mu A_\mu\right]
\end{equation}

where $m$ and $q$ are the particle mass and charge, and $u^{\mu}$ is the four-velocity. One can show that if we constrain the definition for the four-potential to be invariant under the augmentation, then:

\begin{equation}
m \dfrac{\text{d} v_\alpha}{\text{d} t} = - q \mu_0 \Pi_{\alpha \beta} \dfrac{\text{d} x^{\beta}}{\text{d} t}
\end{equation}

The augmentation does perturb $\Pi_{\alpha \beta}$, however the force does no work; it is in fact a fictious force, and therefore can be omitted. Finally, we omit the spatially averaged current to avoid a build up of loop voltage; one can show that if we take the spatially averaged part of the Maxwell-Ampere law:

\[
- \int v(f_{\text{ion,0}} + f_{\text{tur,0}}) \text{d} v = \partial_t E_0 + \alpha_0 E_0
\]

We require that the mean current is very small, and is dominated by exponential decay. In such a case, we find that the spatially averaged electic field $E_0(t)$ must be temporally evanescent:

\[
E_0(t) \approx E_0(t=0) e^{-\alpha_0 t}
\]

This in turn allows us to remove the spatially averaged electric field by setting $E_0(t=0) = 0$ as a boundary condition.

\section{\label{app:ghost} Seed electric field}
If one chooses the following form for the seed contribution:

\[
\mathcal{S}_j \equiv - \dfrac{1}{2}\sum\limits_s \left[A_{js} e^{-i \omega_s t} +  \text{c.c.} \right] 
\]

where $\{A_{js}\} \in \mathbb{C}$, one finds the following partial differential equation:

\[
\partial_t E_{\mathcal{S},j} + \dfrac{1}{2}\alpha_j E_{\mathcal{S},j} = \dfrac{1}{2}\sum\limits_s \left[A_{js} e^{-i \omega_s t} +  \text{c.c.} \right] 
\]

We define Laplace forwards and backwards transforms by:

\[
\begin{array}{r c l}
\tilde{f}(p) \equiv \displaystyle\int\limits_0^{\infty} f(t) e^{-pt} \dif t &;& f(t) \equiv \dfrac{1}{2 \pi i} \lim\limits_{T \to \infty} \displaystyle\int\limits_{\sigma - i T}^{\sigma + i T} \tilde{f}(p) e^{pt} \dif p
\end{array}
\]

This allows one to find that under forward Laplace transformations:

\[
\tilde{E}_{\mathcal{S},j} = \dfrac{1}{p+\alpha_j/2} \left\{ E_{\mathcal{S},j}(t=0) + \sum\limits_s \left[\dfrac{A_{js}}{p + i \omega_s} + \dfrac{A_{js}^*}{p + i \omega_s} \right]  \right\}
\]

Formally, for convergence of the forward transform:

\[
\exists \sigma < \text{Re}(p): \lim_{t \to \infty} |f(t)| = e^{\sigma t} 
\]

Therefore, if the traditional Bromwich contour is shifted to examine a line integral along $\text{Re}(p) < \sigma$, we can examine singularities in $\tilde{E}_{\mathcal{S},j}$; residues of these singularities allow us to recover the solution for the electric field via the residue theorem. By examining $\text{Re}(p) \to -\infty$, we find that the only remaining contribution to the backwards transform is the residues, as the rest of the integral becomes exponentially small:

\[
E_{\mathcal{S},j} = 2 \pi i \cdot \displaystyle\sum\limits_r \text{Res} (\tilde{E}_{\mathcal{S},j} e^{pt}, p_r)
\]

where $p_r$ are the locations of the singularities for $\text{Re}(p) < \sigma$. When calculated, this yields:

\[
E_{\mathcal{S}}(x,t) = E_{\mathcal{S}}^{(\text{ev})}(x) e^{-\alpha_j t/2} + \dfrac{1}{2}\sum\limits_s \left[E_{\mathcal{S},{js}}(x) e^{-i \omega_s t} + \text{c.c.} \right]
\]

where $E_{\mathcal{S}}^{(\text{ev})}$ refers to non-propagating evanescent modes given by the simple pole at $p = - \alpha_j/2$. We can deny these modes from existing by setting $E_{\mathcal{S}}^{(\text{ev})} = 0$ as a boundary condition. $E_{\mathcal{S},{js}}(x)$ is given by:

\[
E_{\mathcal{S},{js}} = \dfrac{A_{js}}{\alpha_j/2 - i \omega_s}
\]

If we select $A_s \in \mathbb{R}$, we find that this reduces to the form:

\[
E_{\mathcal{S}}(x,t) = \sum\limits_s \dfrac{A_s}{\alpha_j^2/4 + \omega_s^2} \cos (\omega_s t)
\]

If one solves the modified Maxwell-Ampere law when there is very little change to the distribution function (negligible instability drive) and no noise, we find the solution:

\[
\delta E(t) = -\epsilon_E E_{\mathcal{S}}(x,t)
\]

As can be seen, for these modes there is no net electric field overall; the seed mode and the perturbation are counterpropagating. Physically, this corresponds to a launched light wave in the plasma being perfectly reflected; note that we have discarded evanescent modes.

We then consider the addition of a noise term $E_{\mathcal{N}}$. By similar analysis, one can show that the noise is representable as a distinct set of frequencies, $\{\omega_n\}$. However, in the limit that $\{\omega_n\}$ form a continuum, we find that we can represent the noise term contribution in the form:

\[
\lim_{\delta \omega_n \to 0}\dfrac{1}{2} \sum\limits_n \left[A_n e^{-i \omega_n t} + \text{c.c.} \right] \equiv - \mathcal{N}_E
\]

where $\mathcal{N}_E$ is a pseudorandom noise term that seeds instabilities. We can once again ignore evanescent effects via boundary conditions, leaving only the propagating contribution. Again, one finds from a similar analysis that the noise term leads to no net perturbation of $E$, in accordance with the conservation of energy.

However, one can show that by solving \eqref{eq:ks4} for negligible current:

\[
\delta E_j (t + \Delta t) \approx \dfrac{\mathcal{D}_j}{\alpha_j/2} \left[1 - \exp\left(- \dfrac{1}{2} \alpha_j \Delta t \right) \right] + \mathcal{O}(\Delta t^2)
\]

The numerical flaw associated with timestep size is what gives the initial drive; clearly, the limit as $\Delta t \to 0$ yields $\delta E_j = 0$.

\section{Computional method}
\subsection{\label{app:strang} Strang splitting}
Formally, the system's equations of motion are defined by a rank two tensor of differential operators $\hat{Z}^{\mu}_{\nu}$, acting in the phase-space $(\mathbf{x},\mathbf{v})$ on the state vector $F^{\mu}$:

\begin{equation}
\partial_t F^{\mu} = \hat{Z}^{\mu}_{\nu} F^{\nu}
\end{equation}

The flow operator tensor $\Phi^{\mu}_{\nu}$ yields the trajectory of the system:

\begin{equation}
F^{\mu}(\mathbf{x}, \mathbf{v}, t + \Delta t) = \Phi^{\mu}_{\nu}(\Delta t) \circ F^{\nu}(\mathbf{x}, \mathbf{v}, t)
\end{equation}

We then suppose that $\hat{Z}^{\mu}_{\nu}$ is representable as a linear sum of $n$ operators. In such a case:

\begin{equation}
Z^{\mu}_{\nu} = \sum\limits_{j=1}^{n} (Z_j)^{\mu}_{\nu}
\end{equation}

Therefore, via the Baker-Campbell-Hausdorff formula [\onlinecite{schur1889neue}], one finds that as for two matrices $\underline{\mathbf{A}}$ and $\underline{\mathbf{B}}$:

\begin{equation}
e^{(\mathbf{A} + \mathbf{B})\Delta t} = e^{\mathbf{A} \Delta t} e^{\mathbf{B} \Delta t} + \mathcal{O}(\Delta t)
\end{equation}

the overall flow can be split into $n$ partial flows via the symmetric Strang splitting method [\onlinecite{strang1968construction,cheng1976integration}]:

\begin{equation}
\begin{array}{l}
F^{\mu}(\mathbf{x}, \mathbf{v}, t + \Delta t) = \left[\prod\limits_{j=1}^{n}(\Phi_j)^{\mu}_{\beta}(\Delta t/2) \circ \right] \\ 
\hspace{40pt} \left[\prod\limits_{j=1}^{n}(\Phi_{n-j})^{\beta}_{\nu}(\Delta t/2)  \circ \right] F^{\nu}(\mathbf{x}, \mathbf{v}, t) + \mathcal{O}(\Delta t^2)
\end{array}
\end{equation}

Accordingly, we first split the problem by evolving one of the functions in the state vector while keeping the rest constant. Secondly, we split the flow for this state function into partial flows. Then each partial flow is solved numerically in forward flow order and then reverse flow order, for a half a timestep each.

For example, in the $x-v$ code, we utilise a state vector of $(f, E)$, keeping the electric field constant while the distribution function is evolved, and vice versa. The partial flows examine spatial advection, velocity advection, collisions, and electric field evolution.

\end{appendix}

\end{document}